\documentclass[letterpaper]{article} % DO NOT CHANGE THIS

\usepackage[draft]{aaai25}

\usepackage{times}  % DO NOT CHANGE THIS
\usepackage{helvet}  % DO NOT CHANGE THIS
\usepackage{courier}  % DO NOT CHANGE THIS
\usepackage[hyphens]{url}  % DO NOT CHANGE THIS
\usepackage{graphicx} % DO NOT CHANGE THIS
\usepackage{natbib}  % DO NOT CHANGE THIS AND DO NOT ADD ANY OPTIONS TO IT
\usepackage{caption} % DO NOT CHANGE THIS AND DO NOT ADD ANY OPTIONS TO IT
\usepackage{algorithm}
\usepackage{algorithmic}

\usepackage{newfloat}
\usepackage{listings}
\DeclareCaptionStyle{ruled}{labelfont=normalfont,labelsep=colon,strut=off} % DO NOT CHANGE THIS
\floatstyle{ruled}
\newfloat{listing}{tb}{lst}{}
\floatname{listing}{Listing}
\usepackage{hyperref} %-- This package is specifically forbidden

\title{Computing Game Symmetries and Equilibria That Respect Them\footnote{Long and updated version to the published paper in the Proceedings of the 39th Annual AAAI Conference on Artificial Intelligence (AAAI 2025).}}
\author{
    Emanuel Tewolde\textsuperscript{\rm 1,\rm 2}, Brian Hu Zhang\textsuperscript{\rm 1}, Caspar Oesterheld\textsuperscript{\rm 1,\rm 2},
    \\
    Tuomas Sandholm\textsuperscript{\rm 1,\rm 3,\rm 4,\rm 5}, Vincent Conitzer\textsuperscript{\rm 1,\rm 2}
}
\affiliations{
    \textsuperscript{\rm 1}Carnegie Mellon University, 
    \textsuperscript{\rm 2}Foundations of Cooperative AI Lab (FOCAL),\\
    \textsuperscript{\rm 3}Strategy Robot, Inc., \textsuperscript{\rm 4}Strategic Machine, Inc., \textsuperscript{\rm 5}Optimized Markets, Inc.\\
    \{etewolde, bhzhang, coesterh, sandholm, conitzer\}@cs.cmu.edu
}
\usepackage{import}
\usepackage{personal_preamble}

\begin{document}

\setcounter{page}{1}

\maketitle

\begin{abstract}
Strategic interactions can be represented more concisely, and analyzed and solved more efficiently, if we are aware of the symmetries within the multiagent system. 
Symmetries also have conceptual implications, for example for equilibrium selection.
We study the computational complexity of identifying and using symmetries. Using the classical framework of normal-form games, we consider game symmetries that can be across some or all players and/or actions. We find a strong connection between game symmetries and graph automorphisms, yielding graph automorphism and graph isomorphism completeness results for characterizing the symmetries present in a game. On the other hand, we also show that the problem becomes polynomial-time solvable when we restrict the consideration of actions in one of two ways.

Next, we investigate when exactly game symmetries can be successfully leveraged for Nash equilibrium computation. We show that finding a Nash equilibrium that respects a given set of symmetries is PPAD- and CLS-complete in general-sum and team games respectively---that is, exactly as hard as Brouwer fixed point and gradient descent problems. Finally, we present polynomial-time methods for the special cases where we are aware of a vast number of symmetries, or where the game is two-player zero-sum and we do not even know the symmetries.
\end{abstract}

\section{Introduction}
\label{sec:intro}

\subsection{Motivation}
In AI and decision making, we appreciate the presence of symmetries, and they are of utmost importance in game theory and multiagent systems. For one, central concepts such as cooperation, conflict, and coordination are usually presented most simply on \emph{totally symmetric} games\footnote{To be defined later;  informally, games in which players have the same strategy options and take on the same ``role'' in the game.}, such as the Prisoner's Dilemma, Chicken, and Stag Hunt. The classic and performant Lemke-Howson algorithm for finding \NEs{} is frequently (and without loss of generality) presented for totally symmetric games~\cite[Section~2.3]{NisanTV07} for the sake of clarity. Furthermore, sometimes interactions with symmetries can be described more concisely in comparison to enumerating the full outcome payoff functions: ``Matching Pennies is a two-player game where each player has two actions $\{0,1\}$. If both players play the same action, player 1 wins, otherwise, player 2 wins.'' This is oftentimes leveraged in games where we \emph{design} the outcome and reward structures, such as in social choice \cite{Brandt+15} and mechanism design \cite{Moulin04} via anonymity, neutrality, and fairness axioms.

\begin{figure}[t]
    \centering
    \includegraphics[width=6.5cm]{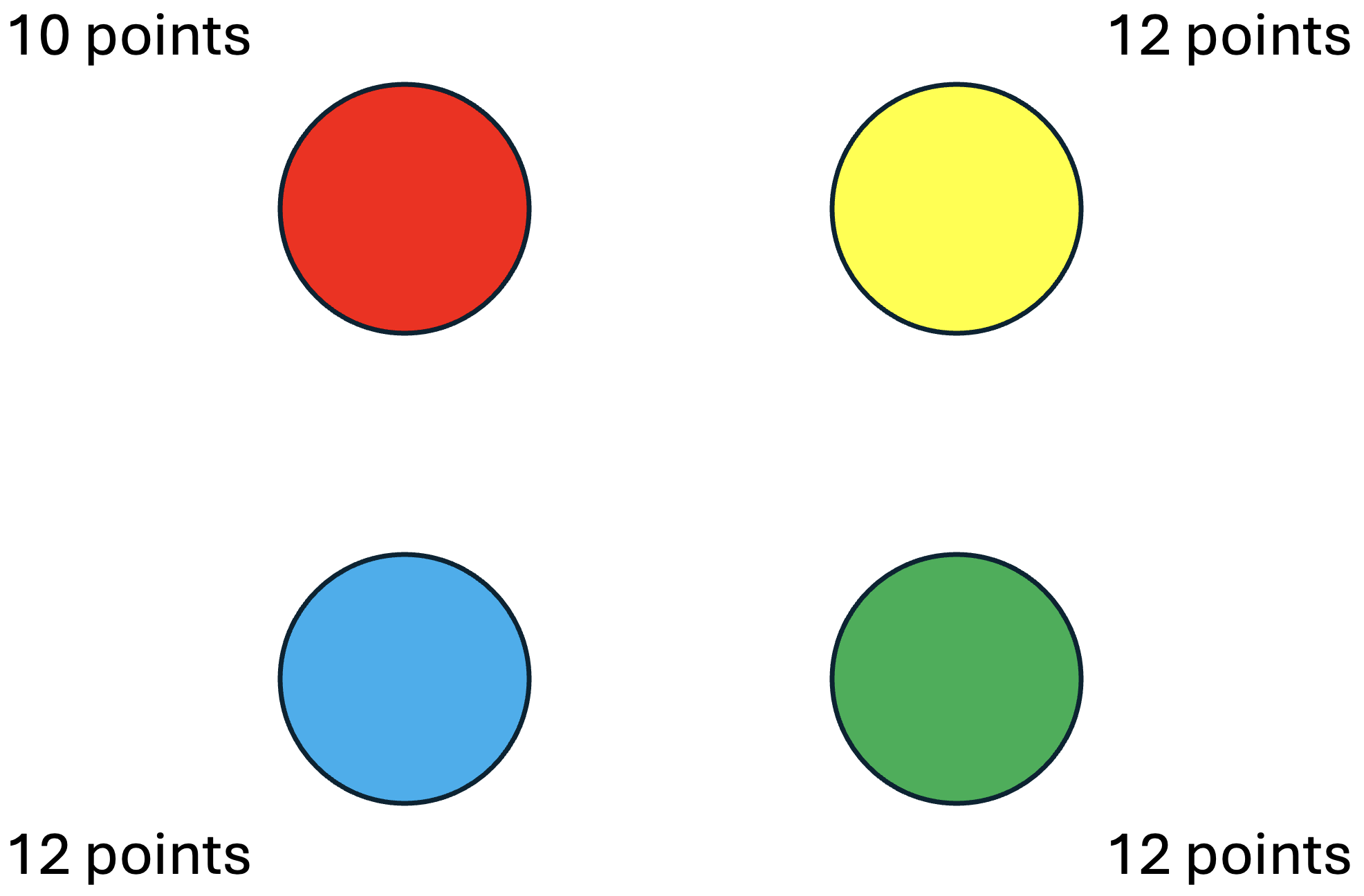}
    \caption{A two-player coordination game. If both players pick the same color, they each receive the associated utility points. If they miscoordinate, both receive $0$ points. Without knowing who you are playing with, what color would you choose?}
    \label{fig:coord}
\end{figure}

Indeed, notions of \emph{fairness} have been connected to the premise that any participant of the game might be assigned to any player identity in the game, which forms a symmetry across participants~\cite{GaleKT52}. For the sake of fairness, one would then like the player identities to be equally strong (\cf the Matching Pennies game, and the ``veil of ignorance'' philosophy~\cite{Rawls71,EmmonsOCC022}). \citet{Ham21}, and the references therein, give a formal treatment of this in terms of game symmetries.

\begin{table*}
\centering
\begin{tabular}{c|cc|cc|c}
    & \multicolumn{2}{c}{Game symmetries} & \multicolumn{2}{c}{Game isomorphisms} & Computing a symmetric equilibrium\\\hline
    General games & \multirow{3}{*}{\makecell{\GA{}-c; \\ (Th. \ref{thm:GA compl})}}& \multirow{3}{*}{\makecell{\XP$\left(\frac{\textnormal{\#actions}}{\textnormal{\#players}} \right)$ \\ (Th. \ref{thm:all symm algo})}}  & \multirow{3}{*}{\makecell{\GI{}-c; \\ (GGS11)}}& \multirow{3}{*}{\makecell{\XP$\left(\frac{\textnormal{\#actions}}{\textnormal{\#players}} \right)$ \\ (Th. \ref{thm:all symm algo})}}   & \PPAD-c (Prop. \ref{thm:PPAD}); \XP($\textnormal{\#orbits}$) (Th. \ref{thm:poly if low num orbits}) \\\cline{1-1} \cline{6-6}
    Team games & & & & & \CLS-c (Th. \ref{thm:CLS})
    \\\cline{1-1} \cline{6-6}
    Zero-sum games &  &  & & & \FP{} (Th. \ref{thm:2p0s NEs easy}) \\\hline
    1PL-actions symmetries & \multicolumn{4}{c|}{\P{} (Prop. \ref{prop:one act set permuting symms poly})} & \multirow{2}{*}{\PPAD-c (Prop. \ref{thm:PPAD})}\\\cline{1-5}
    Player symmetries & \multicolumn{4}{c|}{\P{} (Th. \ref{thm:player symms})}
\end{tabular}
\caption{A high-level summary of some complexity results we obtain across various special cases of games and restrictions on the symmetry sets; though we refer to the associated results for exact statements. `-c' denotes completeness for the respective class. We obtain the hardness results for very narrow settings already, such as, for example, two-player games. \XP($k)$ stands for runtimes in which the only exponent is $k$.}
\label{table:results}
\end{table*}

This symmetry idea that any participant (AIs, humans, etc.) might take on any player identity in the game (\eg, black versus white in chess) also reappears when reasoning about other agents of which we do not have a prior: since the beginnings of machine learning, it has been popular to learn good strategies in \emph{self-play}~\cite{Samuel59}, that is, to assume that other players would use the same strategy as oneself. Self-play continues to be a core contributor to AIs that can learn with no or limited access to human data, and reach super-human performance in domains such as Go~\cite{Silver+16,Silver+17}, and two- and multi-player poker~\cite{BrownS18,BrownS19}. Beyond leveraging the player symmetry in chess and Go by always orienting the board from the moving player’s perspective, \citeauthor{Silver+17} also exploit the rotation and reflection symmetries in Go.

Related to self-play, we may also utilize game symmetries for the purposes of strategy pruning and equilibrium selection~\cite{HarsanyiS88}. Consider the coordination game in~\Cref{fig:coord}. In an ideal scenario, the two players manage to coordinate on the same color between the three that yield the maximal reward of $12$ points. However, if there is no further basis for distinguishing the high-reward colors (\cf focal points~\cite{Schelling60,Alosferrer13}), then
the players run a significant risk of miscoordinating if they attempt to get the reward of $12$. In fact, a natural strategy in this game is to instead go for red with the lower reward of 10. We can explain this formally by recognizing that there are symmetries permuting the colors \{Y,B,G\} for both players while keeping the players' preferences over colors unchanged. Hence, without access to some prior coordination device between the players, the players cannot properly differentiate between \{Y, B, G\}. Therefore, the players should assign the same likelihood of play across those colors, that is, select a strategy profile that respects the aforementioned symmetries. Under this constraint, both players picking red becomes the unique optimal (\NE{}) profile. The equilibrium that uniformly randomizes over \{Y,B,G\} merely achieves an expected reward of $4$.\footnote{With a coordination device, the players are able to achieve a reward of $12$ while respecting the symmetries, namely, by uniformly randomizing over profiles \{(Y, Y), (B,B), (G,G)\}. Correlated equilibria, however, will not be a focus in this paper.} In more recent work, \citet{HuLPF20} and \citet{TreutleinDOF21} apply this argument to zero-shot coordination problems in order to tackle the shortcomings of standard self-play. We remark that respecting the color and player symmetries does not hurt the players if they play this game repeatedly instead. In that case, they can achieve a long-term average of $12$ points by, for example, both playing the following symmetric strategy: In round $1$, randomize uniformly over \{Y, B, G\}. In round $t\geq2$, repeat last round's action if both of you coordinated successfully last round. Otherwise, repeat last round's action only with $50\%$ chance, and the other player's action from last round with the other $50\%$ chance.

Some strategic interactions may also force symmetric play across different decision points. This could be, for instance, because multiple agents (say, self-driving cars) run the same software for taking decisions. In another example, an agent may not recall being in the same situation before (\emph{absentmindedness}) because, \eg, the agent does not retain any record of its history. In this case, it will necessarily act in the same fashion as it did before. \Cref{thm:CLS} makes precise and exploits that there does not appear to be a sharp distinction between (1) being absentminded, (2) being multiple copies of the same agent, and (3) symmetric agents playing symmetries-respecting profiles. We illustrate the underlying argument for it in \Cref{app:sim and ir} on a variant of an example that \citeauthor{KovarikSHC25} [\citeyear{KovarikOC23}, \citeyear{KovarikSHC25}] give for games with simulation of other players. A broader discussion of the game-theoretic aspects of these AI settings can be found in \cite{Conitzer19,ConitzerO23}.

Last but not least, several methods for finding solutions to multiagent problems make great use of symmetries or awareness thereof. On the applied side of solvers, \citet{MarrisG0TLT22} learn to compute Nash, correlated, and coarse correlated equilibria, and achieve sample efficiency by imposing game symmetry invariance onto their neural network architecture. \citet{LiuMPGH24} extends this to transformer-based representation learning of normal-form games, with which they show state-of-the-art performance on various additional tasks such as predicting deviation incentives. Earlier work~\cite{GilpinT07} has developed an abstraction algorithm for solving large-scale extensive-form games that is based on detecting game symmetries (or a related notion thereof) and merging subgames accordingly. On the theory side, \citet{FabrikantPT04} give a polytime algorithm for pure \NE{} network congestion game whenever all players are symmetric, and \citet{DaskalakisP07} develop a polytime approximation scheme for two-action \emph{anonymous games}---a popular game class with particular kinds of symmetries.

A further discussion of related work can be found in \Cref{app:related}. 

\subsection{Structure and a First Overview}

In \Cref{sec:prelim symm}, we start with background on game symmetries in normal-form games. Our general notion of symmetry encompasses any permutation of players \textit{and} their action sets while keeping the utility payoffs unchanged. This is important: the players in Matching Pennies take on different roles in the game (matcher vs.\ mismatcher), and as such, can only be considered symmetric if we allow swapping the two actions of one player while simultaneously swapping the player identities. In another example, the symmetries discussed for the coordination game of \Cref{fig:coord} keep player identities the same and only permute the action sets. In \Cref{sec:comp symm} we connect the presence of symmetries in a game to the presence of symmetries in a graph, and vice versa. The latter is a well-studied computational problem from which we obtain some of the complexity results summarized in \Cref{table:results}. Not included in this table are \Cref{thm:turing eqvl} and \Cref{thm:not player sep,thm:conj resolved}. They focus on characterizing the \emph{set} of game symmetries and relate it to the graph isomorphism problem. As a consequence, \Cref{thm:conj resolved} resolves an open conjecture by \citet{CaoY18} on deciding whether a game is \emph{name-irrelevant symmetric}. Furthermore, our proof ideas can also be applied to the related \emph{game isomorphism} problem, and so we simultaneously discuss those implications.

\Cref{sec:NE prelim} introduces \NEs{} that respects a given set of symmetries, then relates it to group-theoretic ideas involving orbits of actions, and further discusses computational preliminaries. In \Cref{sec:eq search}, we present a series of results on the complexities of computing \NEs{} that respect a given set of symmetries or all symmetries. A summary can again be found in \Cref{table:results}. We give a contextualized discussion of these results in the upcoming \Cref{sec:discuss}, and accompany it with additional insights.

Full proofs can be found in the appendix.

\subsection{Are Symmetries Actually Helpful for Solving Games?}
\label{sec:discuss}

As discussed in the introduction, symmetries have been successfully used for state-of-the-art equilibrium computation methods. Nonetheless, we should not be too quick to conclude that symmetries, if present, ought to be used. 

\paragraph{Potential Harm} We have already seen that in the coordination game of \Cref{fig:coord}, the players might actually strongly prefer to play \NEs{} that fail to respect symmetries of the game. This effect is amplified in that game if we take away the color red from the alternatives, leaving us with a maximal symmetry-respecting payoff of $4$. However, we also note here that a similar argument can be given for the opposite position. Take the totally symmetric two-player game of chicken, that is, the bimatrix game $(A, A^T)$ where $A = \begin{pmatrix} 0 & -1 \\ 1 & -10 \end{pmatrix}$. In a \NE{} that respects the symmetry that swaps the players, the players play their first strategy with probability $0.9$, yielding each of them a payoff of $0.1$. If they instead each search for an asymmetric \NE{}, they may find distinct equilibria---for example, perhaps
each player find the equilibrium that is best for that player. This results in the two players miscoordinating, resulting in the worst of all outcomes (each playing their second strategy). 

\paragraph{Potential Slow-down} Players also might want to ignore symmetries present in a game for the sake of faster computation. Take a totally symmetric bimatrix game $\Gamma = (A, A^T)$ with payoffs in $[0,1]$. It has long been known~\cite{GaleKT52} that finding a symmetric \NE{} of such a game cannot be easier than finding any \NE{} of a general bimatrix game, which makes it \PPAD{}-hard. Now consider the symmetric bimatrix game $(\tilde A, \tilde A^T)$ defined by $\tilde A = \begin{pmatrix} -10 & 2 \cdot \1^T \\ 2 \cdot \1 & A \end{pmatrix}$, 
where $\1$ denotes the vector of appropriate dimension with all entries $=1$. This game has its Pareto-optimal \NEs{} located at strategy profiles that are obvious to find\footnote{Concretely, it requires parsing the full payoff matrix while recognizing that payoffs are in $[0,1]$ in A. This takes linear time.} for any participant: one player must play their first strategy and the other player any strategy but their first. Yet, if we restrict ourselves to respect the player symmetry, then both players playing the first strategy suddenly becomes unattractive. It leaving us with no choice but to find a \NE{} of the original $\Gamma$, which is a \PPAD{}-hard task. This phenomenon becomes even more omnipresent in team games---also known as identical-interest or common-payoff games---because such games are guaranteed to have Pareto-optimal \NEs{} in a pure strategy profile. However, these profiles might not respect most or any nontrivial symmetries present in the game, as illustrated in the coordination game of~\Cref{fig:coord}. Instead, the constraint of respecting symmetries leaves us with the harder computational problem of non-linear continuous optimization, as we will show in \Cref{thm:CLS}. 

\paragraph{Results and General Conclusions} This goes to show that for computational efficiency as well as for achieving high payoffs, one might want to be informed about the game before imposing  the constraint of respecting symmetries. This stands in contrast to some self-play approaches---such as when using regret learning with full feedback---which implicitly respect symmetries, and other solving techniques mentioned in the introduction that have symmetry explicitly imposed into their architecture. 

In \Cref{thm:PPAD}, on the other hand, we show that the requirement of respecting a given set of symmetries does not make the search for a \NE{} harder in the worst-case (\PPAD{}-completeness), and in \Cref{thm:CLS} we show that gradient descent methods are the best we can generally do in team games if we want a given set of symmetries to be respected (\CLS{}-completeness). An additional special case that arises is with the class of two-player zero-sum games (\Cref{thm:2p0s NEs easy}): without having to compute any symmetry of the game (which we show to be graph automorphism hard), we can find a \NE{} that respects \textit{all} of the game's symmetries in polytime via a convex optimization approach.

\begin{figure} 
    \begin{minipage}{.49\linewidth}
    \centering
        \begin{subfigure}{\textwidth}
        \centering
        \includegraphics[width=3.8cm]{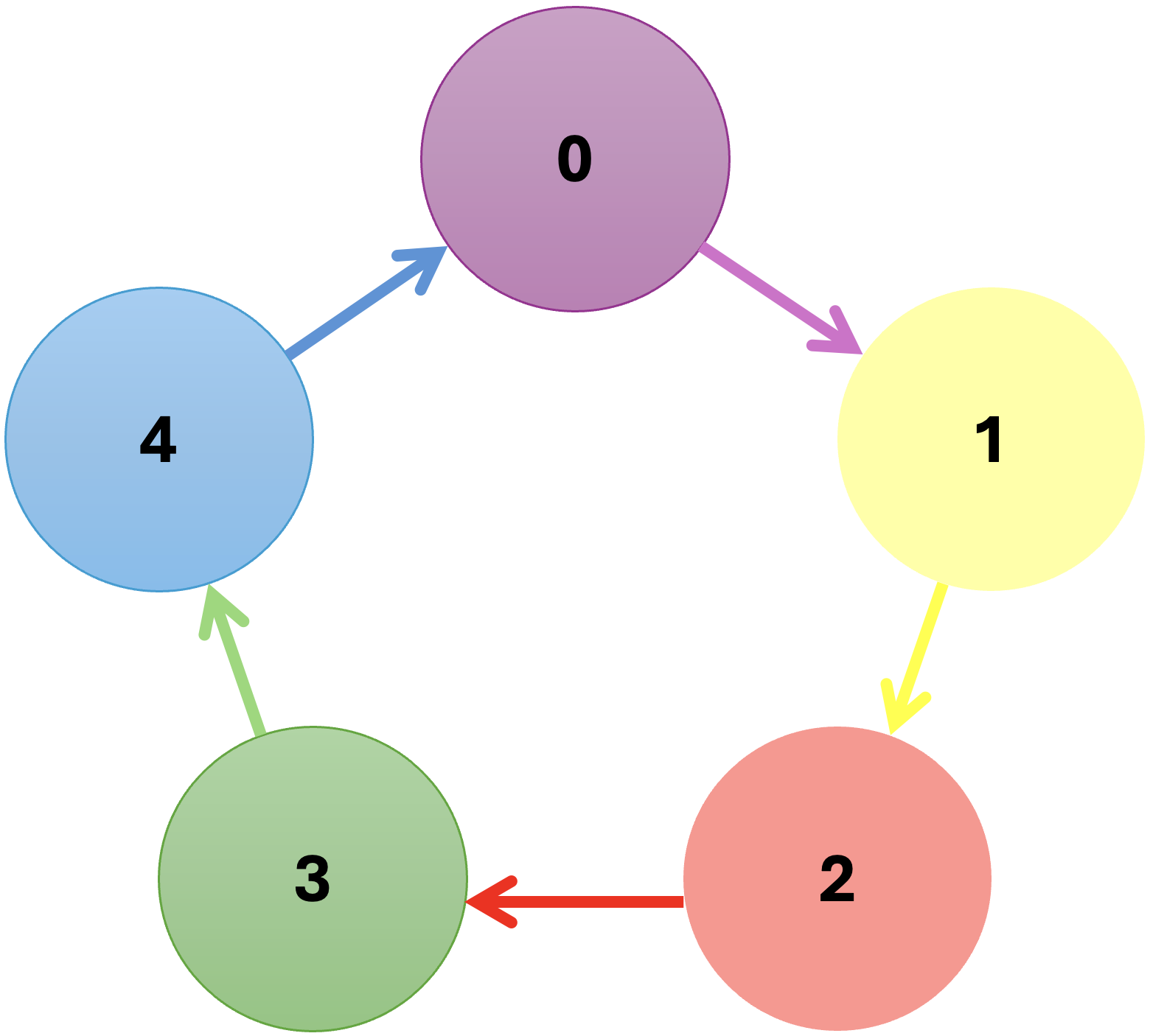}
        \end{subfigure}
    \end{minipage}
    \begin{minipage}{.49\linewidth}
    \centering
        \begin{subfigure}{\textwidth}
        \centering
        \includegraphics[width=3.8cm]{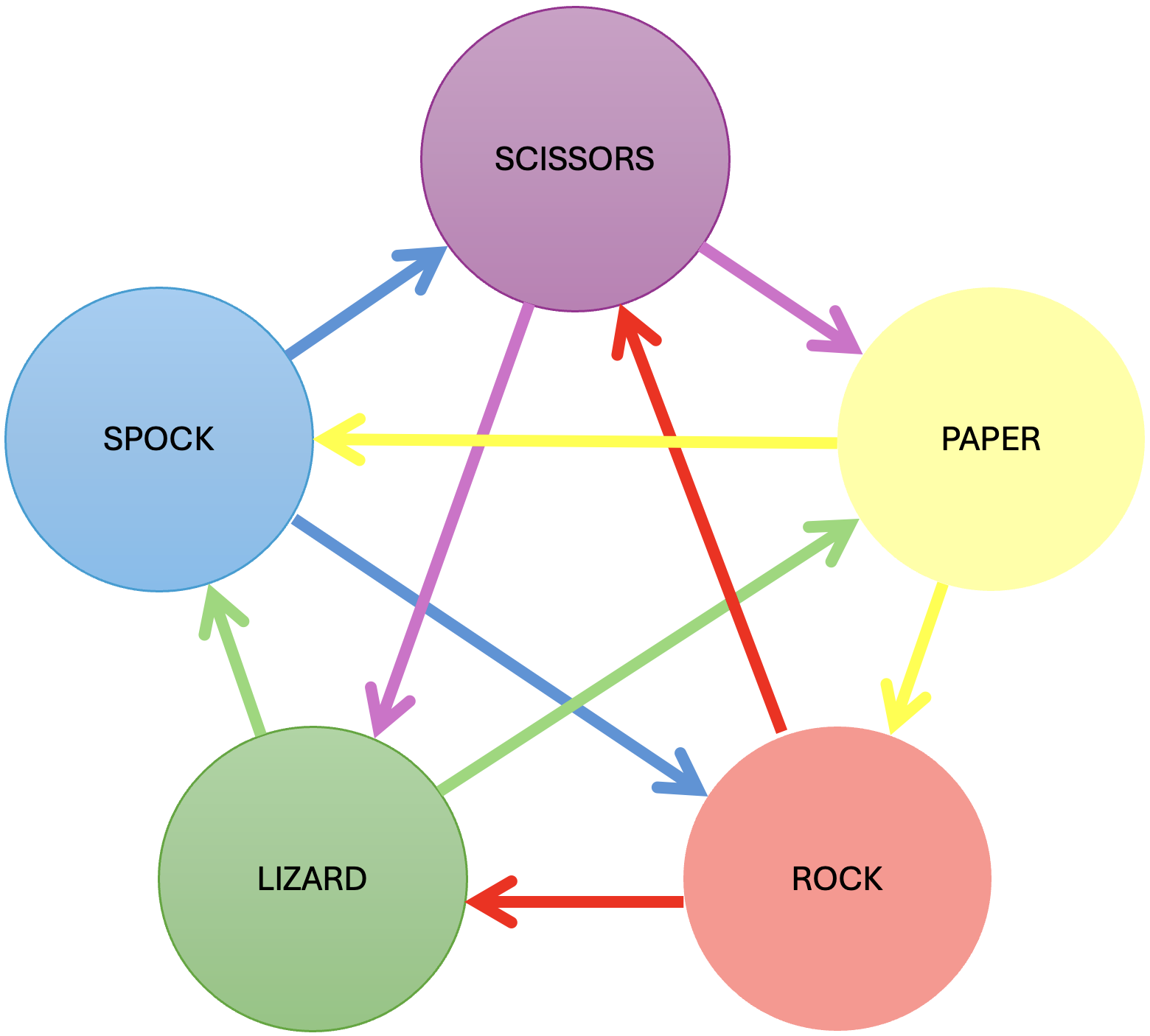}
        \end{subfigure}
    \end{minipage}
    \caption{Two extensions to Rock-Paper-Scissors. In both, there is only one symmetries-respecting strategy profile. The right game is known as Rock-Paper-Scissors-Lizard-Spock. }
    \label{fig:rps}
\end{figure}

\paragraph{A Positive Result When There Are Many Symmetries} With many players in the game, the normal-form representation blows up exponentially, casting that representation impractical. That is why many-player games are usually represented more concisely, often making use of a vast number of symmetries present in the game. So what can we say then? If we are aware of enough symmetries between players and actions such that we are left with a constant number of \emph{orbits} of actions, we can compute a \NE{} that respects those symmetries in polytime (\Cref{thm:poly if low num orbits}). This generalizes a result by \citet{PapadimitriouR05}, and we can illustrate it on an $N$-player $m$-action variant of Rock-Paper-Scissors: each player $i$ chooses an action $j$ from $\{0, \dots, m-1\}$, upon which they receive a payoff equal to ``\# wins - \# losses'' where \# wins is the total number of players in the game choosing action $j + 1 \, (\textnormal{mod } m)$ and \# losses the number of players choosing $j - 1 \, (\textnormal{mod } m)$. \Cref{fig:rps} illustrates when an action wins in this game for $m=5$, and it shows another variant that is prominently referenced in pop culture \cite{rpsx}. It is not hard to see that neither of these games change if we rotate the player identities by $1 \to \dots \to i \to \dots \to N \to 1$, or, instead, if we rotate the action labels by $1 \to \dots \to j \to \dots \to m \to 1$ for all players simultaneously. With those symmetries there is only a \emph{single} orbit of actions. This renders the task of finding a symmetry-respecting \NE{} not only polytime but trivial, because there is only one strategy profile left that respects those symmetries: each player uniformly randomizing over all of their action alternatives.

\paragraph{A Remark On Approximate Symmetries}
The symmetry notion we study in this paper requires payoff profiles to match exactly. Earlier in this section, we argued that this is common in real-world scenarios; in particular, when they are human-designed. Furthermore, we believe that our results generalize meaningfully to settings in which it is unlikely to find exact payoff matches, \eg, because utilities are drawn from a random distribution. To illustrate, let us revisit the color coordination game in \Cref{fig:coord}, except now, coordinating on \{Y,B,G\} yields $11.9$, $12$, and $12.1$ points to both. \citet{vanDamme97} then argues that the slightest uncertainties over payoffs---whether due to exogenous stochasticity or private information---may be reason enough for both players to pick red. We leave it to future work to give a general treatment of approximate notions of symmetries.

\section{Preliminaries on Game Symmetries}
\label{sec:prelim symm}

\begin{defn}
    A \emph{(normal-form) game} $\Gamma$ consists of
    \begin{enumerate}
        \item A finite set of \emph{players} $\calN := \{1, \dots, N\}$, where $N \geq 2$ denotes the number of players,
        \item A finite set of \emph{actions} $A^i := \{1, \dots, m^i\}$ for each player $i \in \calN$, where $m^i$ denotes the number of actions, and
        \item A \emph{utility payoff} function $u^i: A^1 \times \dots \times A^N \to \R$ for each player $i \in \calN$.
    \end{enumerate}
\end{defn}

The players' goal is to maximize their own utility. An \emph{action profile} $\bfa$ specifies what action each player takes and the set $A$ denotes the set of all action profiles, that is, $\bfa = (\bfa_1, \dots, \bfa_N) \in A^1 \times \dots \times A^N =: A$. We also denote the set of actions as $\acts := \sqcup_{i \in \calN} A^i$.

\begin{rem}
    For computational considerations, $u^i$ is restricted to evaluate as rational values only. We assume a game $\Gamma$ is given in explicit form, that is, it is stated as $(\calN, (A^i)_{i \in \calN}, (T^i)_{i \in \calN})$, where $T^i$ is a look-up table of length $|A|$ with all of player $i$'s payoffs under each $\bfa \in A$.
\end{rem}
$T^i$ represents an $N$-dimensional payoff tensor; for example, for $N=2$, that is a matrix. By abuse of notation, we also use cardinality $|\cdot|$ to denote the encoding size of an object that is not a set, \eg, $|\Gamma|$ for a game $\Gamma$.

\begin{defn}[\citealp{Nash51}]
\label{defn:game sym}
    A \emph{(game) symmetry} of a game $\Gamma$ is a bijective map $\phi: \acts \to \acts$ that additionally satisfies
    \begin{enumerate}
        \item actions of the same player are mapped to the same player, \ie, for each $i \in \calN$, there is $\pi(i) \in \calN$ satisfying 
        \\
        $a,a' \in A^i \implies \phi(a), \phi(a') \in A^{\pi(i)}$
        \item[2.] payoffs are symmetry-invariant, that is, 
        \\
        $u^i(\bfa) = u^{\pi(i)}(\phi(\bfa))$ for all $i \in \calN$ and $\bfa \in A$.
    \end{enumerate}
\end{defn}

To explain the notation $\phi(\bfa)$, we first remark that the map $\phi$ induces a bijective player map $\pi: \calN \to \calN$ and bijective action set maps $\phi^i := \phi|_{A^i} : A^i \to A^{\pi(i)}$. Map $\pi$ is henceforth referred to as a \emph{player permutation}. By abuse of notation, $\phi$ can then be considered to map an action profile $\bfa \in A$ to action profile $\phi(\bfa) := \big( \phi(\bfa_{\pi^{-1}(j)} ) \big)_{j \in \calN} \in A$.\footnote{We cannot simply define $\phi(\bfa)$ as $(\phi(\bfa_i))_{i \in \calN}$ because the $j$-th action in this vector is an action that belongs to player $\pi(j)$.}

Let us revisit the bimatrix game $(A,B)$ Matching Pennies, where $A = \begin{pmatrix} 1 & -1 \\ -1 & 1 \end{pmatrix}$ and $B = \begin{pmatrix} -1 & 1 \\ 1 & -1 \end{pmatrix}$. If PL1's and PL2's actions are \{up,down\} and \{left,right\} respectively, then this game has a symmetry ``up $\to$ left $\to$ down $\to$ right $\to$ up''. For instance, PL1 receives under profile (up,right) the same as PL2 under (up,left). Compare this with the another popular but more restrictive definition:
\begin{defn}[\citealp{NeumannM44}]
\label{defn:total symm}
    A game $\Gamma$ is called \emph{totally symmetric} if each player has the same action set $A^*$, and if for any player permutation $\pi$ we have $u^i(\bfa) = u^{\pi(i)} \Big( (\bfa_{\pi^{-1}(j)} )_{j \in \calN} \big)$ for any player $i$ and action profile $\bfa \in \times_{i \in \calN} A^*$.
\end{defn}
In bimatrix games, this reduces to $B = A^T$. In particular, Matching Pennies has symmetries, but it is not totally symmetric. 

Until \Cref{sec:NE prelim}, we assume that a game symmetry $\phi$ is represented in \emph{explicit form}, which, again, means as a look-up table of evaluations of $\phi$. Unlike with games, this explicit representation does not blow up exponentially with the number of players since $\phi$ only has $|\acts|$ evaluations.

Note that the identity map $\id_{\acts}$ is always a game symmetry (henceforth the \emph{trivial symmetry}), and that two symmetries $\phi$ and $\phi'$ compose to a third symmetry. Moreover, if $\phi$ is a symmetry, then $\phi^{-1}$ is one as well. Therefore:
\begin{rem}
\label{rem:sym group}
    The set $\sym(\Gamma)$ of symmetries of a game $\Gamma$ forms a group together with map composition.
\end{rem}
Symmetry groups can be exponentially large; up to $N! \cdot \prod_{i=1}^N (m^i!)$ in our case. For purposes of algorithms, we will thus consider a group $G$ as specified by a subset $Z$ of generators\footnote{Set $Z \subset G$ generates a finite group $G$ if any $g \in G$ can be written as a composition of finitely many elements in $Z$.}; writing $\gen{Z} = G$. Every finite group has a generator set of logarithmic size, which is $\log_2(|\acts|!) = \tilde \calO(|\acts|)$ for us.

Last but not least, we define the graph problems that become relevant later on. A {\em simple graph} (hereafter just {\em graph}) $G = (V, E)$ consists of a set of vertices $V$ and a set of edges $E \subseteq \binom{V}{2}$. The encoding size parameters are $|V|$ and $|E|$.
\begin{defn}[\GA{}]
\label{defn:graph auto}
    In the graph automorphism problem \GA{}, we are given a graph $G$, and asked whether $G$ admits a nontrivial \emph{automorphism}, that is, a bijective map $\phi : V \to V$ that is not the identity function, and that satisfies 
    \\
    $(v,w) \in E \implies (\phi(v), \phi(w)) \in E$.
\end{defn}

\begin{defn}[\GI{}]
\label{defn:graph iso}
    In the graph isomorphism problem \GI{}, we are given two graphs $G = (V,E)$ and $G' = (V',E')$, and asked whether there exists any \emph{isomorphism} $G \to G'$, that is, a bijective map $\phi : V \to V'$ that satisfies 
    \\
    $(v,w) \in E \iff (\phi(v), \phi(w)) \in E'$.
\end{defn}

We call a decision problem \GA{}- or \GI{}-complete if it polytime reduces to \GA{} (resp. \GI{}) and vice versa. No polynomial-time algorithms for \GA{} or \GI{} are known, and \GI{} is widely conjectured to be neither in \P{} nor \NP-hard.\footnote{For example, \GI{} being \NP-complete would imply that the polynomial hierarchy collapses~\cite{schoning1988graph}.} \GA{} many-one reduces to \GI{}~\cite{lozano1992non}, but the reverse reduction is unknown. The best algorithm for \GI{} is due to \citet{babai2016graph}, and runs in time $\exp\relax(\log^{\calO(1)}(|V|))$.

\section{Computation of Game Symmetries}
\label{sec:comp symm}

In this section we study the complexity of characterizing the symmetries in a game. As a warm-up, we consider two simple cases. We start by observing that a game is totally symmetric if and only if swapping any two players $i,j \in \calN$ forms a game symmetry. This yields the following result.

\begin{restatable}{prop}{totalsymmpoly}
\label{cor:total symm poly}
    We can determine whether a game is totally symmetric, for $A^* = A^1 = \dots = A^N$ with the current action numbering, in polytime $\calO(N^2 \cdot |A|)$.
\end{restatable}

Next, we study a type of symmetry we have not discussed yet. Let us call $\phi$ a \emph{1PL-actions} symmetry of $\Gamma$ if it merely permutes the actions of a single player, \ie, there is $i \in \calN$ such that $\phi|_{\acts \setminus A^i} = \id_{\acts \setminus A^i}$. We show that those are generated by symmetries $\phi'$ that expose action duplicates, \ie, that swap two actions $a,a' \in A^i$ for a player $i$ and keep the rest fixed. There are $\calO( N \cdot \max_i (m^i)^2)$ such symmetries.

\begin{restatable}{prop}{onePLactspoly}
\label{prop:one act set permuting symms poly}
    We can compute (a generator set of) the group of 1PL-actions symmetries of a game $\Gamma$ in polytime $\calO( N \cdot \max_i (m^i)^2 \cdot |A| )$.
\end{restatable}

\subsection{Complexity Results}
\label{sec:symmcompl}

We note that none of the symmetry examples from \Cref{sec:intro} are 1PL-actions symmetries. Instead, the \{Y,B,G\} symmetries described for the coordination game of \Cref{fig:coord} is what we call \emph{player-separable} because they keep player identities fixed, \ie, a symmetry $\phi$ whose $\pi = \id_{\calN}$. We will show that such symmetries are already hard to characterize, let alone the whole set of symmetries $\sym(\Gamma)$---which, as we recall, allows an arbitrary permutation of players and their action sets simultaneously.

\begin{restatable}{thm}{thmGAc}
\label{thm:GA compl}
    It is \GA{}-complete to decide whether a game has a nontrivial symmetry. Hardness already holds for two-player \{zero-sum / team\} games that only possess game symmetries that are player-separable.
\end{restatable}

The brackets indicate that the hardness works for the zero-sum restriction, but it also works for the team restriction.

\begin{proof}[Proof Idea]
    For membership, create an \emph{edge-labeled} graph with node set $\{ (i,a) : i \in \calN, a \in A^i \} \cup \{ (a, \bfa): a \in \bfa \in A \} \cup \{(i,\bfa): i \in \calN, \bfa \in A\}$. The first kind and second kind of edges shall receive two distinct labels, and edges $(i, \bfa)$ are labeled with $u^i(\bfa)$. Finally, we note that \GA{} remains its complexity when the graph has edge labels. For hardness, create a two-player game with one action per vertex. Next, we give the players different payoffs depending on whether they play the same, neighboring, or non-neighboring vertices. To remove symmetries across players, we can give PL2 an additional dummy action.
\end{proof}

Our proof method carries over to a known \GI{}-completeness result of the related \emph{game isomorphism problem}. This problem is defined similarly to \Cref{defn:game sym}, except now we are given two games $\Gamma$ and $\Gamma'$ and are asking for a player- and utility-preserving map $\phi: \acts \to \acts'$; see \Cref{app:prelim symm} for a formal definition. In particular, a game symmetry is simply a game isomorphism from a game to itself.

\begin{restatable}[Improved from \citealp{GabarroGS11}, Thm.~6]{thm}{thmGabarro}
\label{thm:gabarro griso}
    It is \GI{}-complete to decide whether two games are isomorphic. Hardness already holds for two-player \{zero-sum / team\} games that only possess game symmetries that are player-separable.
\end{restatable}

We think this result is worth noting because \citeauthor{GabarroGS11} only establish hardness for mixed-motive $4$-player games, and because they do not describe why their problem reduces to \GI{}. Indeed, they partly accredit ``personal communication'' with another researcher as a reference.

Furthermore, the proof constructions in \Cref{thm:GA compl} additionally imply that the symmetries $\sym(\Gamma)$ and automorphisms $\aut(G)$ of associated game-graph pair $(\Gamma, G)$ are isomorphic in a group-theoretic sense. Therefore, we can inherit further known results about graph automorphisms for our setting \cite{Mathon79}.

\begin{restatable}{prop}{propTuring}
\label{thm:turing eqvl}
    The following problems for a game $\Gamma$ are polynomial-time Turing-equivalent to \GI{}: (a) determining a generator set of $\sym(\Gamma)$, and (b) determining the cardinality of $\sym(\Gamma)$ . Hardness already holds for two-player \{zero-sum / team\} with only player-separable symmetries.
\end{restatable}

With an independent proof idea, we can additionally obtain hardness of deciding whether different players are symmetric to each other.

\begin{restatable}{thm}{thmplayerchange}
\label{thm:not player sep}
    Deciding whether $\Gamma$ has a symmetry $\phi$ that is not player-separable, \ie, that maps at least one player to another player, is \GI{}-complete. Hardness already holds for two-player zero-sum games.
\end{restatable}

With this result, we can also prove an open conjecture by \citet{CaoY18} in the affirmative.

\begin{restatable}{thm}{thmconjres}
\label{thm:conj resolved}
    It is \GI{}-complete to decide whether a game $\Gamma$ is \emph{name-irrelevant symmetric}, that is, whether for all possible player permutations $\pi: \calN \to \calN$ there is symmetry $\phi \in \sym(\Gamma)$ of $\Gamma$ that induces it. Hardness already holds for two-player zero-sum games.
\end{restatable}

\subsection{Efficient Computation}
\label{sec:effsymm}

Next, we study efficient ways to compute the set of symmetries in a game (resp. isomorphisms between two games). For the results below, we require that each player $i$ has $m^i \ge 2$ actions, that is, there is no player with no impact on the game. For the sake of space and presentation, we only present the statements in terms of game symmetries in this main body and defer to \Cref{app:efficient game symms} for the treatment of game isomorphisms. 

\begin{restatable}{thm}{allsymmcomp}
\label{thm:all symm algo}
    We can compute (a generator set of) the group $\sym(\Gamma)$ of symmetries of a game $\Gamma$ in time $2^{\calO(|\acts|)}$.
\end{restatable}

We prove \Cref{thm:all symm algo} by reducing the problem to a {\em hypergraph} automorphism problem over a hypergraph with $|V| = \calO(|\acts|)$ nodes, and then applying the $2^{O(|V|)}$-time algorithm for hypergraph automorphism due to \citet{luks1999hypergraph}. 

\begin{restatable}{cor}{boundedactionspoly}
\label{cor:bounded actions poly}
    For games in which the number of actions per player is bounded, we can compute $\sym(\Gamma)$ in polytime.
\end{restatable}

This is because then $|\acts| = \calO(N)$, making the algorithm of \Cref{thm:all symm algo} run in time $2^{\calO(N)}$, which is is polytime in the size of the payoff tensors of the game. It gives a new perspective on the \GA{}- and \GI{}-hardness results we proved so far, since they hold even for a bounded ($=2$) number of players: the computational hardness arises from a growing number of actions of multiple players simultaneously.

We further utilize the reduction idea to hypergraph automorphism for games with player symmetries: A game symmetry $\phi$ is called a \emph{player symmetry} if it keeps the action labels ``fixed'', that is, if it sends action $k \in \{1, \dots, m^i\}$ of player $i$ to action $k$ of player $\pi(i)$.

\begin{restatable}{thm}{playersymmcomp}
\label{thm:player symms}
    We can compute (a generator set of) the group of player symmetries of a game $\Gamma$ in polytime.
\end{restatable} 

For this proof, we must be particularly careful that the number of nodes in the constructed hypergraph does not increase linearly with the size of an individual player's action set. This is accomplished by creating $\calO(\log m^i)$ action nodes for each player (instead of $m^i$ as in \Cref{thm:GA compl}), and associating to each action a {\em subset} of these nodes. 

The positive results of \Cref{cor:bounded actions poly} and \Cref{thm:player symms} rely on the fact that the game is given in explicit form. In other, more concise game representations, we might find that these computational problems become hard again. In graphical games \cite{Kearns01:Graphical}, for example, we have easy-to-obtain hardness simply because such games are already conveniently {\em represented} as graphs.
\begin{restatable}{prop}{graphicalgames}
    The game automorphism (resp. isomorphism) problem for graphical games is \GA{}- (resp. \GI{}-)hard, even in team games with $2$ actions per player.
\end{restatable}

\section{Preliminaries on Nash Equilibria That Respect Game Symmetries}
\label{sec:NE prelim}

Beyond giving background definitions in this section, we also study how ``respecting'' symmetries relate to action orbits, and what that implies for the the complexity considerations of \NE{} computation. 

\subsection{Strategies, Nash Equilibria, Respecting Symmetries}

As usual, we allow the players to randomize over their actions. That is, they can choose a probability distribution---called \emph{strategy}---over $A^i$. The strategy sets are denoted by $S^i = \Delta(A^i)$. A strategy profile $\bfs$ and the strategy profile set $S$ are defined similarly to their counterpart for actions: $\bfs = (\bfs_1, \dots, \bfs_N) \in S^1 \times \dots \times S^N =: S$. Utilities naturally extend to $S$ by taking the expectation $u^i(\bfs) := \sum_{\bfa \in A} \bfs_1(\bfa_1) \cdot \ldots \cdot \bfs_N(\bfa_N) \cdot u^i(\bfa)$.
For notational convenience, $\bfs_{-i} \in \bigtimes_{j \neq i} S^{j} =: S^{-i}$ abbreviates the strategies that all players are playing but $i$. 

\begin{defn}
\label{defn:NE}
    A strategy profile $\bfs \in S$ is called a \emph{\NE{}} of $\Gamma$ if for all player $i \in \calN$ and all alternative strategies $s \in S^i$ we have $u^i(\bfs) = u^i(\bfs_i, \bfs_{-i}) \geq u^i(s, \bfs_{-i})$.
\end{defn}
That is, every player plays their optimal strategy taken as given what the other players have chosen. It is well-known that any game admits a \NE{} \cite{Nash48}. Next, we will discuss \citeauthor{Nash51}'s follow-up work that further shows that symmetries-respecting \NEs{} always exist.

Working towards that result, we first observe that a game symmetry mapping $\phi$ naturally extends to probability distributions over actions, \ie, strategies. Thus, we can overload notation and write $S \ni \bfs \mapsto \phi(s)$. Symmetry $\phi$ will then also satisfy the invariance $u^i(\bfs) = u^{\pi(i)}(\phi(\bfs))$. Furthermore, we have:
\begin{restatable}{rem}{symmpresnes}
\label{lem:symm preserves NE}
    For any symmetry $\phi$ of $\Gamma$, we have that strategy profile $\bfs \in S$ is a \NE{} if and only if $\phi(\bfs)$ is.
\end{restatable}
As we have argued in the introduction of this paper, game symmetries may indicate what actions
% ---in the absence of any further context on or conventions of the game---
ought to be played with the same likelihood; \cf, \eg, the discussions of \Cref{fig:coord,fig:rps}. Let $\Sigma \subseteq \sym(\Gamma)$ be a particular set of symmetries that we want to respect. This could be the trivial set $\{\id\}$, in which case no symmetries need to be respected, or the full set $\sym(\Gamma)$. This could also be \emph{any} subset of symmetries that are readily available to us for a particular game, for example, because they are immediately exposed from a verbal description of the game.
\begin{defn}
    A strategy profile $\bfs \in S$ is said to \emph{respect the} \emph{symmetries} $\Sigma \subseteq \sym(\Gamma)$ if for all $\phi \in \Sigma$ we have $\phi(\bfs) = \bfs$.
\end{defn}

\begin{thm}[\citealp{Nash51}]
\label{thm:symm NE exists}
    Any game $\Gamma$ admits a \NE{} that respects \emph{all} symmetries $\sym(\Gamma)$. Hence, it admits a \NE{} that respects any particular set $\Sigma \subseteq \sym(\Gamma)$ of symmetries.
\end{thm}

\citeauthor{Nash51} obtains this result via a Brouwer fixed point argument, and the proof contains a nonconstructive analysis of the set of symmetries-respecting strategy profiles. In order to make this proof constructive and computational in \Cref{thm:PPAD}, we introduce action orbits next.

\subsection{Orbits Are All You Need}

If we are interested in respecting a set $\Sigma$ of symmetries, it suffices to know what actions are mapped to another action under some symmetry in the subgroup $\gen{\Sigma} \leq \sym(\Gamma)$ \emph{generated} by $\Sigma$. This is called the \emph{orbit} of an action $a \in \acts$ under group $\gen{\Sigma}$, denoted by $\gen{\Sigma} a := \{ \phi(a) : \phi \in \gen{\Sigma} \} \subseteq \acts$. The orbits $W(\gen{\Sigma}) = \{ \gen{\Sigma} a : a \in \acts \}$ partition the total set of actions $\acts$. We obtain a characterization that has already been noted in prior work that studied player symmetries.

\begin{restatable}[\cf \citealp{EmmonsOCC022}]{lemma}{symmseqvltoorbits}
\label{lem:symms gen and orbits eqvl}
    Profile $\bfs$ respects a set of symmetries $\Sigma \subseteq \sym(\Gamma)$ if and only if it respects $\gen{\Sigma}$ if and only if for all orbits $\omega \in W(\gen{\Sigma})$ and actions $a, a' \in \omega$ of respective players $i,i'$ we have $\bfs_{i} (a) = \bfs_{i'} (a')$.
\end{restatable}

\subsection{Computational Considerations}

In games of three players or more, the only (symmetry-respecting) \NE{} might take on irrational values \cite{Nash51} even though the game payoffs are integers. In order to represent solutions in finite bit length, we allow approximate solutions up to some precision error $\eps > 0$. An $\eps$-\NE{} must then satisfy $u^i(\bfs) \geq u^i(s, \bfs_{-i}) - \eps$ in \Cref{defn:NE}. We want $\eps$ to be `small' relative to the range of utility payoffs, which---by shifting and rescaling \cite{TewoldeC24}---we can \wlogg assume to be $[0,1]$. Then, $\eps$ is given in binary, \ie, we seek algorithms that depend polynomially on $\log(1/\eps)$. 

Since (1) symmetry-respecting \NEs{} always exist, and (2) we can check whether a strategy profile is indeed a \NE{} that respects a given set of symmetries, we enter the complexity theory landscape of total \NP{} search problems when it comes to \emph{finding} such equilibria. Its subclasses are characterized by the proof technique used to show that each problem instance admits a solution. We will be interested in the subclasses \PPAD{} and \CLS{}, which both lie somewhere in between \FP{} and \FNP{} (the direct analogues to \P{} and \NP{} when we deal with search problems). \PPAD{} (``Polynomial Parity Arguments on Directed graphs'', \citealp{Papadimitriou94}) contains the problems where a solution is guaranteed to exist via a fixed-point argument, and \CLS{} (``Continuous Local Search'', \citealp{DaskalakisP11}) is based on gradient dynamics on a compact polyhedral domain always admitting a solution.

Last but not least, we will consider three representation schemes for the symmetries we require to be respected, in decreasing order of explicitness: (1) \emph{Explicit form}: A set $\Sigma \subseteq \sym(\Gamma)$ is given as a list of symmetries, each given in explicit form. (2) \emph{Orbit form}: A partition $W$ of actions $\acts$ into orbits, with the promise that $W$ is induced by an unknown set of symmetries $\Sigma \subseteq \sym(\Gamma)$. (3) No symmetries are specified and we require that the full set $\sym(\Gamma)$ of symmetries of $\Gamma$ is respected. We think the explicit form is the natural first inclination for a computational analysis. We introduce the orbit form for games that have a concise description in verbal form, for example, using phrases such as ``if either player 1 plays A or player 2 plays B, then X happens''. Clearly, computing an equilibrium with the last ``representation scheme'' is hardest: A \NE{} that respects \emph{all} symmetries in particular respects any subset of symmetries (even if given in orbit form). Moreover, computation with the orbit form cannot be easier than with the explicit form because we can obtain the former efficiently from the latter:
\begin{restatable}{lemma}{explicittoorbits}
    \label{lem:explicit to orbit}
    Given symmetries $\Sigma$ in explicit form we can compute the orbits $W(\gen{\Sigma})$ it induces in time $\calO(|\Sigma| \cdot |\acts|)$.
\end{restatable}
\section{Finding Symmetries-Respecting Equilibria}
\label{sec:eq search}

In this section, we analyze how hard it is to find a \NE{} that respects symmetries. 

\subsection{Complexity Results}

To start with the general case, let {\sc Sym-Nash} denote the search problem that takes a game $\Gamma$ in explicit form, a precision parameter $\eps > 0$ in binary, and symmetries of $\Gamma$ in orbit form. Given that, it asks for a strategy profile $\mu$ of $\Gamma$ that respects the said symmetries and that is an $\eps$-\NE{}.

\begin{restatable}{prop}{symppad}
\label{thm:PPAD}
    {\sc Sym-Nash} is \PPAD{}-complete. \PPAD{}-hardness already holds for two-player games, where the symmetries are given in explicit form $\Sigma \subseteq \sym(\Gamma)$, and $\Sigma$ contains \{just / more than\} the identity symmetry.
\end{restatable}

\begin{proof}[Proof Idea]
    \Cref{sec:discuss} discussed the well-known idea for proving hardness. We obtain membership by fitting Nash's function to \citet{EtessamiY10} framework for showing that a Brouwer fixed point problem is in \PPAD{}.
\end{proof}

The membership---which, to the best of our knowledge, forms the novel contribution---shows that we can find symmetries-respecting \NEs{} with fixed-point solvers and path-following methods, just as it is the case with finding \emph{any} \NE{} in a normal-form game. Hence, this is a positive algorithmic result. \citet{GargMVY18} proved a related \FIXP{}$_a$-completeness result for exact computation of a player-symmetric \NE{} in a totally symmetric game of constant number of players.

Next, we narrow down our interest to {\sc Sym-Nash-Team}, which we define as the restriction of {\sc Sym-Nash} to the special case of team games, \ie, games with $u^1 = \ldots = u^N$.

\begin{restatable}{thm}{symcls}
\label{thm:CLS}
    {\sc Sym-Nash-Team} is \CLS{}-complete. \CLS{}-hardness already holds for totally symmetric team games of five players where the player symmetries that show total symmetry are given in explicit form.
\end{restatable}
\begin{proof}[Proof Idea]
    We leverage a strong connection between single-player decision making under imperfect recall and decision making in a team under symmetry constraints \cite{LambertMS19}, inheriting known \CLS{}-hardness results for problems in the former setting \cite{TewoldeOCG23,Tewolde+24}. For membership, we show that symmetries-respecting \NEs{} correspond to first-order stationary points (formally, \textit{Karush-Kuhn-Tucker} (KKT) points) of the following polynomial optimization problem: Maximize the team's utility function over the polyhedral domain of symmetries-respecting strategy profiles. \citet{FearnleyGHS23} have shown that finding an approximate KKT point of such a problem is in \CLS{}.
\end{proof}

The \CLS{}-membership shows that first-order methods are suited to find a symmetries-respecting \NEs{} in team games. This has already been observed for player symmetries (1) by \citet{EmmonsOCC022} for the exact gradient descent dynamics and (2) by \citet{GhoshH24} for the two-player case. Our \CLS{}-hardness result, on the other hand, shows that gradient descent is the most efficient algorithm---modulo polynomial time speedups and barring major complexity theory breakthroughs---that is available for this problem. The main result of \citeauthor{GhoshH24}'s concurrent work shows that \CLS{}-hardness already holds for totally symmetric \emph{two-player} games. We remark that this time the trivial symmetry set $\Sigma = \{\id_{\acts}\}$ does not suffice for hardness, because we can find an arbitrary \NE{} of a team game in linear time by going through the payoff tensor and selecting the payoff-maximizing action profile. This is different in game representations that are not normal-form: \citet{BabichenkoR21}---and to a lesser extent \citet{DaskalakisP11}---study the concisely represented \emph{polymatrix games} and \emph{$c$-polytensor games} for $c \in \N$. They also obtain a \CLS{}-completeness result for the team game case, but this time it is for finding any \NE{}.

\subsection{Efficient Computation}

In view of the hardness results in \Cref{thm:PPAD} and \Cref{thm:CLS}, we may ask why it is so popular to leverage game symmetries in game solvers, as discussed in the introduction. Indeed, restricting our attention to symmetries-respecting strategy profiles does allow for a significant dimensionality reduction in the to-be-studied profile space. Unfortunately, one (or a few) game symmetries do not allow for enough of a reduction to affect the asymptotic computational complexity. However, if the number of symmetries is vast, or equivalently, the number of distinct orbits is low, then we can show that equilibrium computation becomes easier. This generalizes a similarly derived result by \citet{PapadimitriouR05} for games with total symmetry, and we have illustrated that in the discussion of the Rock-Paper-Scissors extensions in \Cref{fig:rps}. Furthermore, the color coordination game in \Cref{fig:coord} reduces to the simple, yet nontrivial polynomial optimization problem $\max_{r,\bar r \geq 0, r + \bar r = 1} \, 10 r^2 + 4 \bar{r}^2$, where variables $r$ and $\bar r$ denote the probabilities assigned to the ``red'' and ``the other'' orbit.

\begin{restatable}{thm}{lownumorbits}
\label{thm:poly if low num orbits}
    {\sc Sym-Nash} can be solved in time $\poly\qty( |\Gamma|, \log(1/\epsilon), (|W|N )^{|W|} )$. In two-player games, this can be improved to exact equilibrium computation in time $\poly\qty( |\Gamma|, 2^{|W|} )$.
\end{restatable}
This runtime is polynomial in the input size whenever the symmetries-induced number of orbits $|W|$ is bounded. 
\begin{proof}[Proof Idea]
    We show that symmetries-respecting strategy profiles correspond one-to-one to ``orbit profiles'' in $\R^{|W|}$, which indicate with what probability a player plays any action in a particular orbit. This lower-dimensional space can be described efficiently, and the \NE{} conditions now make a system of $\calO(|W|)$ additional \emph{polynomial} (in-)equalities, where each polynomial has degree at most $N$. Therefore, we can invoke known algorithms for solving such a system from the \emph{the existential theory of the reals} \cite{renegar1992computational}, which will achieve the desired runtime. If it is a two-player game, we can use {\em support enumeration} \cite{dickhaut1993program} on orbit profiles instead.
\end{proof} 

Lastly, we move our attention to two-player zero-sum games, which can be solved for a \NE{} in polytime \cite{Neumann28,Dantzig51,Adler13}. We establish that we can even ensure that \emph{all} symmetries present in the game are respected {\em without} having to compute all / any symmetries of the game (which we know is as hard as \GI{} and \GA{} due to \Cref{thm:turing eqvl} and \Cref{thm:GA compl}).

\begin{restatable}{thm}{twoplzerosumeasy}
\label{thm:2p0s NEs easy}
    Given a two-player zero-sum game, we can compute an exact \NE{} that respects all symmetries present in the game in polytime.
\end{restatable}
\begin{proof}[Proof Idea]
    The set of \NEs{} of a two-player zero-sum game is a convex polytope that can be described efficiently via a system of linear (in-)equalities (\cf minimax theorem, \citealt{Neumann28}). Hence, we can solve any convex quadratic objective over this domain to exact precision in polytime \cite{Kozlov80}. Intuitively, we recognize that changing a strategy profile towards respecting symmetries equates to increasing its probability entropy. Formally, we analyze the ``symmetric'' regularizer objective $S^1 \times S^2 \to \R, \bfs \mapsto \sum_{a \in \acts} f(\bfs(a))$
    for an arbitrary strictly convex function $f : [0, 1] \to \R$, \eg, $x \mapsto x^2$. We prove that a \NE{} that minimizes this objective also respects all symmetries present in the game.
\end{proof}

We believe this result can generalize to other representations or solution concepts, as long as the solution space is an efficiently describable convex compact polytope.
\section{Conclusion}

The concept of symmetry is rich, with many applications across the sciences, and in AI in particular. For game theory, the situation is no different. Indeed, a typical course in game theory conveys the most basic concept of a symmetric (two-player) game; to check whether it applies, no more needs to be done than taking the transpose of a matrix.  But there are other, significantly richer symmetry concepts as well, ones that require relabeling players' actions or which do not allow arbitrary players to be swapped. We have studied these richer concepts.  First, we studied the problem of {\em identifying} symmetries in games, and exhibited close connections to graph iso- and automorphism problems. We also devised performant algorithms for this task, and discussed special cases that have polytime guarantees. Second, we studied the problem of computing solutions to games that {\em respect} their symmetries. We have shown that requiring to respect them does not worsen the algorithmic complexity (significantly), and that it improves the complexity when the number of symmetries is vast. We also gave a strongly positive result for two-player zero-sum games.

There are many directions for future research, including the following.  (1) We have focused on normal-form games (and briefly, graphical games). What about other ways to represent games, such as extensive-form games, stochastic games, and compact representations such as action-graph games~\cite{Jiang11:Action} and MAIDs~\cite{Koller03:Multi}? (2) We have focused on exact symmetries; what about approximate symmetries and other informative relations between players and strategies? (3) There are many conceptual questions regarding symmetries. For example, in many games, the players would benefit from being able to break the symmetries, such as in the color coordination game in the introduction, or from adopting distinct roles (say, on a soccer team).  What are effective and robust ways to break symmetries to achieve better outcomes?

\appendix

\section*{Acknowledgements}
We are grateful to the anonymous reviewers for their valuable improvement suggestions for this paper. Emanuel Tewolde, Caspar Oesterheld, and Vincent Conitzer thank the Cooperative AI Foundation, Polaris Ventures (formerly the Center for Emerging Risk Research) and Jaan Tallinn’s donor-advised fund at Founders Pledge for financial support. Caspar Oesterheld's work is also supported by an FLI PhD Fellowship. Brian Hu Zhang’s work is supported in part by the CMU Computer Science Department Hans Berliner PhD Student Fellowship. Tuomas Sandholm is supported by the Vannevar Bush Faculty Fellowship ONR N00014-23-1-2876, National Science Foundation grants RI-2312342 and RI-1901403, ARO award W911NF2210266, and NIH award A240108S001.

\bibliography{aaai25}

\clearpage

\section{More Proof Idea Sketches}
\label{app:proofideas}

Here, we give some additional proof idea sketches that are missing in the main body due to space constraints.

\begin{proof}[Proof Idea of \Cref{thm:not player sep}]

    Construct the graph $G$ with node set $\calN \sqcup \acts \sqcup A$ as in \Cref{thm:GA compl}, and then check whether $G$ has an isomorphism that maps any player $i \in \calN$ to a different player. We show this to reduce to \GI{}.
    
    We show hardness with an independent proof: Given graphs $G_1$ and $G_2$, create a zero-sum game of two players, each with action set $V_1 \cup V_2$. When both players pick vertices from different graphs, nothing happens. In the case where both choose vertices from the same graph $G_i$, player $i$ will strongly prefer to choose a neighbor of the other player's chosen vertex, and gets slightly punished instead if they do not manage to do so.
\end{proof}

\begin{proof}[Proof Idea of \Cref{thm:conj resolved}]
    Hardness follows from \Cref{thm:not player sep}. For inclusion, check whether, for every pair of players $i$ and $j$, the graph $G$ constructed in \Cref{thm:GA compl} has an automorphism $\phi$ whose restriction to $\calN$ is the transposition $(ij)$. We can show that this reduces to \GI{}.
\end{proof}

\section{On Simulation and Imperfect Recall}
\label{app:sim and ir}

Let us describe the connections between (1) being absentminded, (2) being multiple copies of the same agent, and (3) symmetric agents playing symmetries-respecting profiles. We illustrate it on a variant of an example that \citeauthor{KovarikSHC25} [\citeyear{KovarikOC23}, \citeyear{KovarikSHC25}] give for games with simulation of other players: Consider the perspective of an AI agent that handles a user's financial assets.  It can either invest them, giving the agent utility $1$, or steal them, giving the agent utility $3$. However, the user is not na\"ive, and decides to first run the agent {\em in simulation}, but only once.  If the user catches the agent stealing in simulation, the agent will not actually be given control of the assets, and will walk away with utility~$0$.  Crucially, the agent is unable to tell the difference between being run in simulation and being run in the real world. Therefore, it must steal with some probability $p$ whenever called upon to act, and will thus receive expected utility $(1-p)^2 + 3p(1-p)  = -2p^2+p+1$. Hence, it is optimal for the agent to steal with probability $1/4$ for an expected utility of $9/8$. We can capture this decision problem by modeling the agent as being absentminded, since it will not remember having been in the same situation before. Alternatively, we can also model the decision problem as a game between two distinct players that form a team, one acting first in simulation and one acting second in the real world. In order to do so, we shall randomly assign the two roles (simulated or real) to the two players without revealing the assignment to them. This creates a symmetry between the two players. If we then require that the players respect this symmetry, they will both choose to steal with the same probability $p$. Indeed, without the random and secret role assignment, the team can achieve utility $3$ by letting the player in simulation invest and the player in the real world steal. Without the constraint to respect the symmetry, the team prefers to have one player invest and the other player steal, because then  they steal successfully $50\%$ of the time, achieving utility $3/2 > 9/8$. Hence, both modeling steps are needed to accurately reflect the decision problem described.

\section{Other Related Work}
\label{app:related}

In this section, we describe further related work.

First, let us give more context on the open conjecture we resolved in \Cref{thm:conj resolved}: \citet{CaoY18}, more generally, studied total symmetric, \emph{renaming symmetric}, and name-irrelevant symmetric games from a structural lens. A game is called renaming symmetric if one can ``relabel'' the action set of each action set, such that---if one considers them as the same action set $A^*$---the relabeled game is totally symmetric. They derive \GI{}-completeness for deciding whether a two-player team game with \{0,1\} utility payoffs is renaming symmetric.

\citet{ConitzerS08}, \citet{KontogiannisS11}, \citet{GargMVY18}, and \citet{BiloM21} study various decision and approximation problems involving symmetric \NEs{} in totally symmetric games, such as counting or maximizing payoffs and many more. \citet{MehtaVY15} show that it is \NP{}-complete to decide whether a non-symmetric \NE{} exists in a totally symmetric (bimatrix) game. \citet{BrandtFH09} study concisely represented totally symmetric games, anonymous games, and minor variants, in terms of the complexity of finding a pure \NEs{}. They continue this study in follow-up work \cite{BrandtFH11} with a focus on graphical games. \citet{ChengRVW04} show that (possibly asymmetric) pure \NE{} exist in $2$-action totally symmetric games, and how symmetries can boost equilibrium solvers based on Nelder-Mead variants or replicator dynamics methods. Related to another existence result of theirs, \citet{Reny99} shows existence of symmetric \NE{} in games that have particular kinds of player symmetries, and whose action spaces $A^i$ are compact subsets of a topological vector spaces (among other conditions).

\section{Further Background on \Cref{sec:prelim symm}}
\label{app:prelim symm}

In this section, we define the game isomorphism problem, and give more background on the graph isomorphism and automorphism literature.

\subsection{Notation}
Analogous to strategies, we write $\bfa_{-i} \in A^{-i}$ for the parts in $\bfa$ and $A$ that are associated to the other players from $i$.

\subsection{Game Isomorphisms}

\begin{defn}[\citealt{GabarroGS11}]
\label{defn:game iso}
    A \emph{(game) symmetry} between two games $\Gamma = (\calN, (A^i)_{i \in \calN}, (u^i)_{i \in \calN})$ and $\Gamma' = (\calN', (A'^i)_{i \in \calN}, (u'^i)_{i \in \calN})$ is a bijective map $\phi: \acts \to \acts'$ satisfying
    \begin{enumerate}
        \item actions of the same player are mapped to the same player, that is, for each $i \in \calN$, there is $\pi(i) \in \calN'$ with $a,a' \in A^i \implies \phi(a), \phi(a') \in A'^{\pi(i)}$
    \end{enumerate}    
    [In particular, bijective $\phi$ then induces a bijective player map $\pi: \calN \to \calN$ and bijective action set maps $\phi^i := \phi|_{A^i} : A^i \to A'^{\pi(i)}$. In particular, action profiles $\bfa \in A$ are mapped to action profiles $\phi(\bfa) := \big( \phi(\bfa_{\pi^{-1}(j)} ) \big)_{j \in \calN'} \in A'$.
    \begin{enumerate}
        \item[2.] payoffs are invariant under the isomorphism, that is, $u^i(\bfa) = u'^{\pi(i)}(\phi(\bfa))$ for all $i \in \calN$ and $\bfa \in A$.
    \end{enumerate}
\end{defn}

Just as with game symmetries, we assume a game isomorphism is represented in \emph{explicit form}, that is, as a look-up table of evaluations of $\phi$.

\subsection{Graph Isomorphisms and Automorphism}
\label{app:gi and ga background}

First, a comment on why there are indeed generator sets of logarithmic size: If set $Z = \{z_1, \dots, z_k\}$ is an inclusion-wise minimal generating set of group $G$, then $Z$ generates at least $2^k$ distinct elements, namely $z_S := \prod_{i \in S} z_i$ for each subset $S \subseteq [k]$.

An {\em edge-labeled} (hereafter simply {\em labeled}) {\em graph} is a tuple $G = (V, E_1, \dots, E_k)$,  where $E_i$ is the set of edges with label $i$. Two labeled graphs are isomorphic if there is an edge-label-preserving isomorphism between them. In this section, we state some known or simple results about various problems related to edge-labeled graphs.
\begin{prop}\label{apx:gi}
The following decision problems are \GI{}-complete:
\begin{enumerate}
\item The isomorphism problem for edge-labeled graphs
\item Given $k$ pairs of (labeled) graphs $(G_1, H_1), \dots,$ $(G_k, H_k)$, decide whether $G_i \cong H_i$ for {\bf all} $i$ \label{it:gi-all}
    \item Given $k$ pairs of (labeled) graphs $(G_1, H_1), \dots,$ $(G_k, H_k)$, decide whether $G_i \cong H_i$ for {\bf any} $i$ \label{it:gi-any}
    
    (Note that for $k=1$ both of the previous problems are just graph isomorphism for labeled graphs)
    \item Given two (labeled) graphs $G, H$ and pairs $(u_1, v_1), \dots, (u_k, v_k) \in V_G \times V_H$, decide whether there exists an isomorphism $\phi : V_G \to V_H$ such that $\phi(u_i) = v_i$ for all $i$. \label{it:gi-restricted}
\end{enumerate}
\end{prop} 
\begin{proof}
In all cases, hardness follows because \GI{} is a special case. So lets move to membership:
\begin{enumerate}
    \item This is a special case of Theorem 2 of \citet{miller1977graph}, where the edge sets $E_i$ form what they call \emph{relations} $R_i$.
    \item[2.+3.] These problems are known to be in {\GI} for simple graphs~\cite{lozano1992non}. The labeled version then follows by combining with the previous item.
    % \item Same as the previous item.
    \item[4.] Assume that the $u_i$'s are all different and that the $v_j$'s are all different; otherwise, either the instance is trivially false or there are duplicate pairs, which can be removed. Let $G^{u_1, \dots, u_k}$ be the graph $G$ except with a long ``chain'' / ``line'' of nodes (say, of length $\ell_i \ge V(G)$ for $1 \leq i \leq k$, with the $\ell_i$'s all different) of new vertices attached to $u_i$ for each $i$. Then an isomorphism with the desired property exists if and only if $G^u \cong G^v$. \qedhere
\end{enumerate}
\end{proof}
\begin{prop}
\label{prop:labeled GA to GA}
    The automorphism problem for edge-labeled graphs is \GA{}-complete.
\end{prop}
\begin{proof}
    Follows immediately from a construction identical to Theorem 2 of \citet{miller1977graph}. We also note that the association between automorphism in the graph and and automorphisms in the corresponding edge-labeled graph is a group isomorphism itself, hence questions about the cardinality and generator sets of the automorpshism groups correspond as well.
\end{proof}

\section{On \Cref{sec:comp symm} up until \Cref{sec:symmcompl}}

In this section, we provide proofs to the results of \Cref{sec:comp symm} excluding \Cref{sec:effsymm}.

\totalsymmpoly*

\begin{proof}
    First, we can test whether indeed $m^1 = \dots = m^N$ in linear time. Next, observe that a game is totally symmetric iff all player permuations $\pi \in \calS_\calN$ are player symmetries iff all player swaps $\sigma_{ij}$ for $i,j \in \calN$ are player symmetries (since they generate $\calS_\calN$). The latter can be checked in time $\calO(N^2 \cdot |A|)$. 
\end{proof}

\onePLactspoly*
\begin{proof}
    Suppose $\phi$ is a 1PL-actions symmetry for player $i$ that maps action $a \in A^i$ to $a' \in A^i$. Let $\bfa_{-i} \in A^{-i}$ be arbitrary. Then
    \[
        u^i(a, \bfa_{-i}) = u^{\pi(i)}(\phi(a, \bfa_{-i})) = u^{i}(a', \bfa_{-i}) \, .
    \]
    This implies that $a$ and $a'$ are duplicate actions, always achieving the same payoffs. In particular, the map $\phi'$ that exposes the duplicates $a, a'$ is also a symmetry. But then, any 1PL-actions symmetry for player $i$ is identified by how it permutes $A^i$, which is generated by the swaps of any two actions in $A^i$ (transpositions). Of those swaps, there are $N \cdot \max_i (m^i)^2$ many. Checking if a single swap makes a symmetry takes $\calO(|A|)$ time.
\end{proof}

\thmGAc*

\begin{proof} \,
    \paragraph{Membership} 
    Given a game $\Gamma$, create a labeled graph $G$ with node set $\calN \cup \acts \cup A$. Let $\lambda \neq \lambda' \in \N$ be values that do not occur as payoffs anywhere in $\Gamma$. We want the following edges in the graph: 
    \begin{enumerate}
        \item $\{i,a\}$ with label $\lambda$ for any pair $i \in \calN$ and $a \in A^i$
        \item $\{a, \bfa\}$ with label $\lambda'$ for any pair $\bfa \in A$ and $a \in \bfa$, and
        \item $\{i, \bfa\}$ with label $u^i(\bfa)$ for any pair $i \in \calN$ and $\bfa \in A$.
    \end{enumerate}
    
    We can define a map $M$ that maps a symmetry $\phi$ of $\Gamma$ to an automorphism $M(\phi) := \psi$ of $G$. It is naturally defined by $\psi(i) := \pi(i)$, $\psi(a) := \phi(a)$, and $\psi(\bfa) := \phi(\bfa)$. Then the labeled edges remain invariant under $\psi$, with the only point to note that this association indeed yields for any edge $\{i, \bfa\}$ that $u^i(\bfa) = u^{\pi(i)}(\phi(\bfa))$, and thus $\{\psi(i), \psi(\bfa)\}$ is the same labeled edge in $G$.

    We can show that $M$ is a bijection from symmetries of $\Gamma$ to automorphisms of $G$. Injectivity follows immediately. $M$ is also surjective. Say $\psi$ is a graph automorphism of $G$. We first show that $\psi(\calN) \subseteq \calN$, $\psi(A^i) \subseteq \acts$, and $\psi(A) \subseteq A$, while simultaneously defining the preimage symmetry candidate $\phi$ of $\Gamma$. Let $i \in \calN$. Then for $a \in A^i$ and $\bfa \in A$, we have edges $\{i, a\}$ and $\{i, \bfa\}$. Hence, $\{\psi(i), \psi(a)\}$ and $\{\psi(i), \psi(\bfa)\}$ are also edges in $G$ with the same respective labels $\lambda$ and $u^i(\bfa)$. The only vertices in $G$ that has neighbors as $\psi(i)$ are the ones in $\calN$, hence $\psi(i) \in \calN$. Hence, also $\psi(a) \in A^{\psi(i)} \subset \acts$, and $\psi(\bfa) \in A$. Since $\psi$ is a bijection, we get that $\psi$ induces bijections $\pi := \calN \to \calN$, $\phi := \acts \to \acts$, and $\phi|_{A^i} = A^i \to A^{\pi(i)}$. Through edges $\{a, \bfa\}$ and $\{\psi(a), \psi(\bfa)\}$, we also obtain that if $a \in A^i$, then $A^{\pi(i)} \ni \phi(a) = \psi(a) \in \psi(\bfa)$. Hence, $\psi(\bfa)$ is equal to $\phi(\bfa) := \big( \phi(\bfa_{\pi^{-1}(j)} ) \big)_{j \in \calN} \in A$. It is clear that $\phi$ is indeed the preimage of $psi$ under $M$, \ie, $M(\phi) = \psi$. In order to complete showing that $\phi$ indeed makes a symmetry, it remains to highlight that for $i$ and $\bfa$, we have $u^i(\bfa) = u^{\pi(i)}(\phi(\bfa))$ because $\{i, \bfa\}$ and $\{\psi(i), \{\psi(\bfa)\} = \{\pi(i), \{\phi(\bfa)\}$ have the same edge labels.
    
    Overall, we have that $M$ is a bijective, and we observe that we have correspondence $M(\id_{\acts}) = \id_{V}$ for the identity symmetry and identity automorphism. Using \Cref{prop:labeled GA to GA}, we therefore get \GA{}-membership.

    \paragraph{Hardness}
    Let $G = (V,E)$ be a graph. Create a game $\Gamma$ of two players. PL1's action set shall be $V$, and PL2's action set $V \cup \{ \alpha \}$, where action $0$ is added as a dummy action so that all symmetries of $\Gamma$ will have to keep player identities fixed, as we will see later on. In terms of utility payoffs, we have quite some flexibility: PL1 will receive one of four distinct utility payoffs $\{c_k\}_{k = 1}^4 \subset \R$, and PL2 will receive one of four distinct utility payoffs $\{d_k\}_{k = 1}^4 \subset \R$. To give specific examples, if we want $\Gamma$ to be a team game, we might choose $c_k = k = d_k$ for all $k$; if we want $\Gamma$ to be to be a zero-sum game, we might choose $c_k = k = - d_k$ for all $k$. The players receive payoffs as follows:
    \begin{enumerate}
        \item[(1)] a payoff of $c_1$ and $d_1$ respectively if PL2 chooses action $\alpha$,
        \item[(2)] a payoff of $c_2$ and $d_2$ respectively if the players choose distinct vertices $u,v \in V$ that are not neighbouring ($\{u,v\} \notin E$),
        \item[(3)] a payoff of $c_3$ and $d_4$ respectively if the players choose distinct vertices $u,v \in V$ that are neighbouring ($\{u,v\} \in E$), and
        \item[(4)] a payoff of $c_4$ and $d_4$ respectively if the players choose the same vertex $v \in V$.
    \end{enumerate}

    We can define a map $M$ that maps a symmetry $\psi$ of $G$ to a symmetry $M(\psi) := \phi$ that maps player $i$ to $i$, action $v \in V$ to $\psi(v)$ for both players, and action $\alpha$ to $\alpha$ for PL2. Then each case of payoffs are mapped to themselves: With cases (1) and (4) this is by definition, with cases (2) and (3) this follows from vertices being neighbors if and only if they are after applying $\psi$. So $M(\psi)$ is a game symmetry of $\Gamma$.

    We can again show that $M$ is a bijection. Injectivity again follows immediately. Let us consider surjectivity. We claim that if $\phi$ is a symmetry of the game $\Gamma$, then $\phi|_{A^1}$ is an automorphism of $G$ such that $M(\psi^1) = \phi$. So let $\phi$ be a symmetry of $\Gamma$. Then:
    \begin{enumerate}
        \item The players have action sets of different cardinality. This eliminates the possibility that $\pi(1) = 2$ because in that case $\phi|_{A^1}$ would have to be a bijection between the action sets. Thus, $\pi$ Thus, $\pi$ maps $i$ to $i$.
        \item In particular, $\phi|_{A^1}$ is therefore a permutation of the vertex set $V$.
        \item The players have to receive a utility of $c_1$ and $d_1$ again when PL2 plays $\phi(\alpha)$. This would only be the case for action $\alpha$, thus $\phi(\alpha) = \alpha$.
        \item The players have to receive a utility of $c_4$ and $d_4$ again when, for any $v \in V$, the players play $\phi|_{A^1}(v)$ and $\phi|_{A^2}(v)$. This implies $\phi|_{A^1}(v) = \phi|_{A^2}(v)$ in $V$.
        \item The players have to receive a utility of $c_3$ and $d_3$ (resp. $c_2$ and $d_2$) again when, for any distinct $u,v \in V$ with $\{u,v\} \in E$ (resp. $\{u,v\} \notin E$), the players play $\phi|_{A^1}(u)$ and $\phi|_{A^2}(v) = \phi|_{A^1}(v)$. Therefore, we must have $\{\phi|_{A^1}(u),\phi|_{A^1}(v)\} \in E$ (resp. $\{\phi|_{A^1}(u),\phi|_{A^1}(v)\} \notin E$) in that case as well.
    \end{enumerate}
    Together, this implies that $\phi|_{A^1}$ is a symmetry of $G$, and by definition of $M$, $M(\phi|_{A^1}) = \psi$. 
    
    Overall, we have that $M$ is a bijective, and we observe that we have correspondence $M(\id_{V}) = \id_{\acts}$ for the identity automorphism and identity symmetry. This yields \GA{}-hardness.
\end{proof}

\thmGabarro*
\begin{proof}
    We can use the same proofs as for \Cref{thm:GA compl}. Except now, we start with two games $\Gamma, \Gamma'$ (resp. graphs $G, G'$), and construct each of their corresponding graphs $G, G'$ (resp. games $\Gamma, \Gamma'$). Then, by similar reasoning to the previous proof, we have that the latter pair is isomorphic if and only if the former pair is. In the membership proof, we need to make sure that we use the same $\lambda, \lambda'$ in $G$ and $G'$ and they are different from any payoffs in $G$ and $G'$. In the hardness proof, we need to choose the same values $\{c_k, d_k\}_{k = 1}^4$ for both $\Gamma$ and $\Gamma'$.
\end{proof}

\propTuring*

\begin{proof}
    Taking another look at the proof of \Cref{thm:GA compl}, we have that $M$ commutes with composition of game symmetries and graph automorphisms, that is, $M(\phi' \circ \phi) = M(\phi') \circ M(\phi)$ in the membership proof and $M(\psi' \circ \psi) = M(\psi') \circ M(\psi)$ in the hardness proof. This together with the previously observed characteristics of $M$, implies both $M$'s are group isomorphisms between the symmetry group $\sym(\Gamma)$ and the automorphism group $\aut(G)$ of corresponding $\Gamma$ and $G$ in these two proofs. Therefore, they have equal cardinality and generator sets. Next, transition to the same problems for standard (unlabeled) graphs with \Cref{prop:labeled GA to GA}. Those are polynomial-time Turing-equivalent to \GI{} by \citet{Mathon79}.
\end{proof}

\thmplayerchange*
\begin{proof} \,
    \paragraph{Membership} Take the same graph construction $G$ as in \Cref{thm:GA compl}. We will have $N^2$ instances of the graph isomorphism problem variant where we ask whether $G$ is isomorphic to $G$ via mapping $i \in V$ to $j \in V$, for any $i,j \in \calN \subset V$. From \Cref{apx:gi}, we have that each of these questions reduce to \GI{}, and thus also, that it reduces to \GI{} to decide whether any are true.
     
    \paragraph{Hardness} 
    Let $G_1 = (V_1,E_1)$ and $G_2 = (V_2,E_2)$ be two graphs for which we would like to decide whether they are isomorphic. Construct a two-player zero-sum game $\Gamma$ with action sets $A^1 = V_1 \cup V_2 = A^2$ and payoffs
    \begin{enumerate}
        \item[(1)] $u^1(v,v') = 0 = u^2(v,v')$ if either ($v \in V_1$ and $v' \in V_2$) or ($v \in V_2$ and $v' \in V_1$) player,
        \item[(2)] $u^1(v,v') = 2 = - u^2(v,v')$ if $v, v' \in V_1$ and $\{v,v'\} \in E_1$,
        \item[(3)] $u^1(v,v') = -1 = - u^2(v,v')$ if $v, v' \in V_1$ and $\{v,v'\} \notin E_1$,
        \item[(4)] $u^1(v,v') = -2 = - u^2(v,v')$ if $v, v' \in V_2$ and $\{v,v'\} \in E_2$,
        \item[(5)] $u^1(v,v') = 1 = - u^2(v,v')$ if $v, v' \in V_2$ and $\{v,v'\} \notin E_2$.
     \end{enumerate}

    So in the case where both choose vertices are from the same graph $G_i$, player $i$ will strongly prefer to choose a neighbor of the other player's chosen vertex, and gets slightly punished instead if they do not manage to do so. Nonetheless, these exact effects are not crucial to this construction; intuitively, it is only important that the players receive neighboring-dependent payoffs that are reversed across the two graphs. 

    Let $\psi$ be an isomorphism $G_1 \to G_2$. Then $\Gamma$ has the player swapping symmetry $\pi: 1 \leftrightarrow 2$ and $\phi|_{A^1}(v) := \phi|_{A^1}(v) := \begin{cases} \psi(v) &\text{if } v \in V_1 \\ \psi^{-1}(v) &\text{if } v \in V_2 \end{cases}$. Then this makes a game symmetry. This follows immediately for case (1), and it follows for the other cases, because $\psi$ exactly preserves neighboring vertices, and because cases (2) and (4) as well as cases (3) and (5) have swapped payoffs.

    Let $\phi$ be a symmetry of $\Gamma$ that is not player-separable, that is, where $\pi$ swaps PL1 and PL2. We claim that then $\phi$ maps $V_1 \subset A^1$ to $V_2 \subset A^2$, and that $\psi := \phi|_{V_1}$ is the graph isomorphism $G_1 \to G_2$. Let $v \in V_1$. Then, taking an arbitrary $v' \in V_1$, we have that
    \begin{align}
    \label{eq:not player sep}
        u^2(\phi(v), \phi(v')) = u^1(v,v') = \begin{cases} 2 &\text{if } \{v,v'\} \in E_1 \\ -1 &\text{if } \{v,v'\} \notin E_1 \end{cases} \, .
    \end{align}
    PL2 can only receives those payoffs when $\phi(v), \phi(v') \in V_2$. Analogous reasoning yields that $\phi(V_2) \subseteq V_1$, and hence, by bijectivity of $\phi|_{A^1}$, we have that $\phi|_{V_1}$ bijectively maps $V_1$ to $V_2$. Moreover, above equations (\ref{eq:not player sep}) also yields for $v,v' \in V_1$ that 
    \begin{align*}
        \{v,v'\} \in E_1 &\iff u^1(v,v') = 2 \iff u^2(\phi(v),\phi(v')) = 2 
        \\
        &\iff \{\phi(v),\phi(v')\} \in E_2 \, .
    \end{align*} 
    So $\psi = \phi|_{V_1}$ is a group homomorphism.
\end{proof}

\thmconjres*
\begin{proof}
    Hardness for two-player zero-sum games follows from \Cref{thm:not player sep}, where the only nontrivial player permutation is swapping PL1 with PL2. For membership, we first observe the following fact: Each player permutation $\pi$ is induced by some symmetry $\phi$ of $\Gamma$ if and only if all the transpositions $(i,j)$ for $i,j \in \calN$ as player permutations are induced by some symmetries of $\Gamma$. The forward direction is immediate, and the backward direction uses that any player permutation can be constructed by composing transpositions appropriately (and hence, it suffices to compose the underlying symmetries appropriately to induce a desired player permutation). With that fact, we can conclude the proof with analogous steps to the membership proof of \Cref{thm:not player sep}; except now, we invoke the part of \Cref{apx:gi} that allows us to ask whether multiple graph isomorphism questions are \emph{all} true. Each individual graph isomorphism question, indexed by $(i,j) \in \calN$ will now be as follows: If $G$ and $G'$ are two identical copies of the graph construction of $\Gamma$ as in \Cref{thm:GA compl}, check whether they are isomorphic to each other under the constraint that $i \in G$ is mapped to $j \in G'$, $j \in G$ is mapped to $i \in G'$, and every other player node in $G$ is mapped to itself in $G'$.
\end{proof}

\section{On \Cref{sec:effsymm}}
\label{app:efficient game symms}

In this section, we establish \Cref{thm:all symm algo,thm:player symms}, and also develop their counterpart results for game isomorphisms between two games in \Cref{thm:player iso} and \Cref{thm:iso comp}. Throughout this section, we will assume $m^i \ge 2$ for all players. 

Let $\calS_\calN$ denote the \emph{symmetry group} of permutations on a finite set $\calN$. Later on, $\calN$ will stand for the player set. We begin by specifying a few lemmas about computation in general symmetry groups and cosets. For the purpose of discussing algorithms, subgroups $G \le \calS_\calN$ are specified by generating sets of size $\poly(N)$, and cosets $xG = \{ xg : g \in G\}$ are specified as pairs $(x,G)$. Later on, $x$ will represent a game isomorphism from game $\Gamma$ to game $\Gamma'$. 

First, we will present a few known results. We start with the observation that we can efficiently shrink a generator set to $\poly(N)$ size, and therefore do not need to worry about algorithms that produce larger generator sets.

\begin{lemma}[\citealp{sims1978some}]
Let $Z \subseteq \calS_\calN$. There exists a $\poly(N, |Z|)$-time algorithm that outputs a set $Z' \subseteq Z$ with $|Z'| \le \log_2(N!)$ and $\gen{Z'} = \gen{Z}$.
\end{lemma}

Next, we turn our attention to computing coset intersections sufficiently fast. Say $G, H \le \calS_\calN$ and $x, y \in \calS_\calN$. Then it is not hard to show that there exists an element $z \in xG \cap yH$ if and only if $xG \cap yH=z(G \cap H)$. Hence, in order to compute a coset intersection, it suffices to determine one element $z$ in the coset intersection and a generator set for the group intersection $G \cap H$.
\begin{lemma}[\citealp{luks1999hypergraph}]\label{lem:coset intersection}
    Let $G, H \le \calS_\calN$, and $x, y \in \calS_\calN$. There exists a $2^{\calO(N)}$-time algorithm that computes the coset intersection $xG \cap yH$.\footnote{Faster algorithms exist, \eg, $\exp\relax(\log^{\calO(1)}(N))$ due to \citet{babai2016graph}, but this will be enough for our purposes.}
\end{lemma}

Next, we recall the concept of a \emph{player symmetry}. A permutation $\pi \in \calS_\calN$ acts on game $\Gamma$ by permuting its players according to $\pi$ and leaving action numbering fixed. This does not yet necessarily make a game symmetry of $\Gamma$ if, \eg, some action sets are not equally sized or the utility payoffs do not match. If it does, however, we call it a player symmetry. It preserves utility payoffs in the sense that
\[
    u^i(\bfa) = u^{\pi(i)}(\phi_\pi(\bfa)) = u^{\pi(i)}(\bfa_{\pi^{-1}(1)}, \dots, \bfa_{\pi^{-1}(N)})
\]
for all $i \in \calN$ and $\bfa \in A$. The set of player symmetries form a subgroup of $\calS_\calN$, and is denoted as $\sym_\calN(\Gamma)$.

For efficient encoding, we will work with hypergraphs. A {\em hypergraph} is a pair $G = (V, E)$ of vertices $V$ and \emph{hyperedges} $E \subseteq 2^V$. The notion of an automorphism extends to hypergraphs: It is a bijective mapping $\pi: V \to V$ such that if $V \supseteq W \in E$, then $\pi(W) \in E$.

\begin{thm}[\citealt{luks1999hypergraph}]\label{th:hypergraph}
    The automorphism group for a given hypergraph $G$ can be computed in time $2^{\calO(|V|)}$. 
\end{thm}

Note that this bound is tight up to the constant hidden by the big-O because $G$ might have $|E| = 2^{\Theta(|V|)}$ many edges that are included in its encoding in the first place. Next, we show that we can efficiently compute the group of player symmetries $\sym_\calN(\Gamma)$.

\playersymmcomp*

\begin{proof}
    Consider the following hypergraph. There are several types of nodes:
    \begin{itemize}
        \item Create one node $i$ for each player $i \in \calN$.
        \item Create one node $b_{ik}$ for each $k=1, 2, \dots, k_i := \lceil \log_2(m^i) \rceil$. The actions of player $i$ will be, for this proof, identified with subsets of $B_i := \{ b_{i1}, \dots, b_{ik_i} \}$.
        \item Let $P := \{ u^i(\bfa) : i \in \calN, \bfa \in A\}$ be the set of unique payoff values across all players. Create one node $u_{k}$ for each $k = 1, 2, \dots, \lceil \log_2|P| \rceil$. As before, the unique utility values $u \in P$ will be identified with subsets of $U := \{ u_k \}_k$.
    \end{itemize}
    Now for each player $i \in \calN$ and strategy profile $\bfa \in A$, create a hyperedge containing the node of player $i$, the subset of $B_j$ identified with action $\bfa_j$ for every player $j$, and the subset of $U$ identified with the utility value $u^i(\bfa)$. Let $V := \calN \sqcup B_1 \sqcup \dots \sqcup B_N \cup U$ be the vertex set of this hypergraph.
    
    Let $G \le \calS_V$ be the automorphism group of the hypergraph. Furthermore, define the subgroup $H \le \calS_V$ of vertex permutations as $\{ \psi_\pi \}_{\pi \in H_\calN}$, where (1) the player permutation subgroup $H_\calN \leq \calS_N$ contains exactly those $\pi$ that satisfy $m^i = m^{\pi(i)}$, \ie, we map players with the same action set sizes to each other, and where (2) vertex permutation $\psi_\pi$ for $\pi \in H_\calN$ maps $\calN \ni i \mapsto \pi(i)$, $b_{ik} \mapsto b_{\pi(i)k}$, and $u_k \mapsto u_k$.

    \paragraph{Observation 1:} $\{ \psi_{\pi} : \pi \in \sym_\calN(\Gamma) \} = G \cap H$. The set inclusion ``$\subseteq$'' is clear, since any player symmetry $\pi$ of $\Gamma$ indeed induces an automorphism of the form $\psi_\pi$ of this hypergraph construction.  ``$\supseteq$'': Let $\psi \in G \cap H$, that is, $\psi = \psi_\pi$ for some $\pi \in H_\calN$ and $\psi_\pi$ is a hypergraph automorphism. Then for any $i \in \calN$ and $\bfa \in A$, since there is a hyperedge to $(i, \bfa, u^i(\bfa))$ in $E$, there will also be the hyperedge $(\pi(i), \psi_\pi(\bfa), u^i(\bfa))$ in $E$. This translates to $u^{\pi(i)}(\bfa_{\pi^{-1}(1)}, \dots, \bfa_{\pi^{-1}(N)}) = u^i(\bfa)$. Hence, $\pi$ indeed forms a player symmetry, that is, $\pi \in \sym_\calN(\Gamma) \le H_\calN$.
    \paragraph{Observation 2:} Above observation implies that a generator set of $G \cap H$, restricted to $\calN \subset V$, forms a generator set of $\sym_\calN(\Gamma)$.
    
    \paragraph{Observation 3:} Intersection $G \cap H$ can be computed in polytime. To start, we can compute $G$ with \Cref{th:hypergraph}. To that end, we observe that 
    \begin{align*}
        |V| &\leq N + \sum_{i=1}^N \log_2 m^i + N + \log_2 |P| + 1
        \\&= \calO\qty(\sum_{i=1}^N \log_2 m^i) = \calO\qty( \log_2 \prod_{i=1}^N m^i) 
        \\
        &= \calO\qty( \log_2 |A|) \, ,
    \end{align*}
    which yields a running time of 
    \[
        2^{\calO(|V|)} = 2^{\calO\qty(\log_2 |A|)} = \poly(|A|) \, .
    \]
    
    Next, we note that $H$ is generated by the vertex permutations $\psi_{\sigma_{ij}}$ where $i, j \in \calN$ are players with equal action set sizes $m^i = m^j$, and $\sigma_{ij} \in \calS_\calN$ is the player permutation that swaps $i$ with $j$ and keeps the rest fixed. This generator set can be computed in time $\calO\qty(N^2 \cdot |V| ) = \poly(|A|)$. Finally, \Cref{lem:coset intersection} for $x=y=\id \in \calS_V$ allows us to compute the intersection $G \cap H$ in time $2^{\calO(|V|)} = \poly(|A|)$. 
\end{proof}

How does this generalize to game isomorphisms? First, note that we can apply any player permutation $\pi \in \calS_\calN$ to a game $\Gamma$ to get a new game $\pi(\Gamma)$ which has reordered players according to $\pi$, and the action sets are permuted accordingly. By construction, $\pi$ forms a game isomorphism between $\Gamma$ and $\pi(\Gamma)$. Next, if we have two arbitrary games $\Gamma$ and $\Gamma'$ with the same number of players, and upon fixing a numbering of the players, we can \wlogg identify the players sets as the same $\calN = \{1, \dots, N\}$. With this, we define the set of \emph{player isomorphisms} $\iso_\calN(\Gamma, \Gamma') \leq \calS_\calN$ as the set of player permutations for which $\pi(\Gamma) = \Gamma'$. Any player isomorphism in $\iso_\calN(\Gamma, \Gamma')$ then naturally induces a game isomorphism between $\Gamma$ and $\Gamma'$. Upon also fixing a numbering of the player's action sets, a player isomorphism $\pi$ is characterized by satisfying for all $i \in \calN$ and $\bfa \in A$:
\[
    u^i(\bfa) = (u')^{\pi(i)}(\phi_\pi(\bfa)) = (u')^{\pi(i)}(\bfa_{\pi^{-1}(1)}, \dots, \bfa_{\pi^{-1}(N)}) \, .
\]
Notice that when $\Gamma = \Gamma'$, then $\iso_\calN(\Gamma, \Gamma') = \sym_\calN(\Gamma)$.

\begin{lemma}
\label{lem:player iso coset or empty}
    For any two games $\Gamma, \Gamma'$, either $\iso_\calN(\Gamma, \Gamma')$ is empty, or it is a left coset $\iso_\calN(\Gamma, \Gamma') = \pi \sym_\calN(\Gamma)$ for $\pi \in \iso_\calN(\Gamma, \Gamma')$ chosen arbitrarily.
\end{lemma}

\begin{proof}
    Suppose there is a player isomorphism $\pi \in \iso_\calN(\Gamma, \Gamma')$. Then we show that $\iso_\calN(\Gamma, \Gamma') = \pi \sym_\calN(\Gamma)$. If $\sigma \in \sym_\calN(\Gamma)$, then by definition we have  $\pi(\sigma(\Gamma)) = \pi(\Gamma) = \Gamma'$, so $\pi \circ \sigma \in \iso_\calN(\Gamma, \Gamma')$. Conversely, if $\pi' \in \iso_\calN(\Gamma, \Gamma')$ then $\pi^{-1}(\pi'(\Gamma)) = \Gamma$, so $\sigma := \pi^{-1} \circ \pi' \in \sym_\calN(\Gamma)$ satisfies $\pi \circ \sigma = \pi'$.
\end{proof}

With this, we can get a similar result to \Cref{thm:player symms} for game isomorphisms.

\begin{thm}\label{thm:player iso}
    We can compute the set of player isomorphisms $\iso_\calN(\Gamma, \Gamma')$ between two games $\Gamma$ and $\Gamma'$ in polytime.
\end{thm}
\begin{proof}
    By \Cref{lem:player iso coset or empty}, it suffices to determine an element $\pi \in \iso_\calN(\Gamma, \Gamma')$, if existent, and $\sym_\calN(\Gamma)$. We can do the latter in polytime via \Cref{thm:player symms}. So let us study the former.

    Construct a game $\hat\Gamma$ that incorporates both $\Gamma$ and $\Gamma'$ as follows. $\hat\Gamma$ has $2N$ players: the $N$ players originating from $\Gamma$, which we will call $\calN = \{1, 2, \dots, N\}$ as usual; and the $N$ players originating from $\Gamma'$, for which we introduce a separate copy of players $\calN' := \{ 1', 2', \dots, N'\}$. Each player $i \in \hat\calN := \calN \cup \calN'$ has action set $\hat A^i := A^i \cup \{ \bot \}$, where $\bot$ is a special symbol. The utilities are as follows. Let $u^\bot$ be any number that does not appear as the utility of any player in $\Gamma$ or $\Gamma'$, and let $\bot(\bfa) = \{ i : \bfa_i = \bot\} \subseteq \hat\calN$ be the players that player $\bot$ in $\bfa$. If $\bot(\bfa) \notin \{ \calN, \calN'\}$ then all players get utility $u^\bot$. If $\bot(\bfa) = \calN'$, then players $i \in \calN$ get utility $u^i(\bfa_1, \dots, \bfa_N)$, and players $i' \in \calN'$ get utility $u^\bot$. If $\bot(\bfa) = \calN$, then players $i' \in \calN'$ get utility $(u')^{i'}(\bfa_{1'}, \dots, \bfa_{N'})$, and players $i \in \calN$ get utility $u^\bot$. 

    To analyze this construction, associate to each existent player isomorphism $\pi \in \iso_\calN(\Gamma, \Gamma')$ the player permutation $\tau_{\pi} \in \calS_{\hat \calN}$ defined as $\tau_{\pi}(i) = \pi(i)$ for $i \in \calN \subset \hat \calN$ and $\tau_{\pi}(i') = \pi^{-1}(i')$ for $i' \in \calN' \subset \hat \calN$.
    
    \paragraph{Observation 1:} 
    If $\pi \in \iso_\calN(\Gamma, \Gamma')$, then $\tau_{\pi}$ is a player symmetry of $\hat \Gamma$ with $\tau_{\pi}(\calN) = \calN'$. 
    %that bijectively maps $\calN, \calN' \in \hat \calN$ to each other. 
    Indeed, action set sizes match thanks to $\pi$. Moreover, since $\tau_{\pi} : \calN \overset{\text{bij}}{\leftrightarrow} \calN'$, we have that players who received $u^\bot$ before applying the symmetry will continue to do so after applying the symmetry. Last but not least, players who received a payoff other than $u^\bot$ will continue to receive the same payoff after applying the player symmetry $\tau_{\pi}$ since $\pi$ and $\pi^{-1}$ are player isomorphisms.

    \paragraph{Observation 2:} Any player symmetry $\tau$ of $\hat \Gamma$ must map $\tau(\calN) = \calN$ or $\tau(\calN) = \calN'$. 
    %$\calN \overset{\text{bij}}{\to} \calN$ or $\calN \overset{\text{bij}}{\to} \calN'$. 
    This is because of the case $\bot(\bfa) = \calN$, in which, after applying $\tau$, players $i' \in \calN'$ will again need to receive payoffs $\neq u^\bot$, implying $\tau(\calN) \in \{ \calN, \calN'\}$.

    \paragraph{Observation 3:} If there is a player symmetry $\tau$ of $\hat \Gamma$ mapping $\tau(\calN) = \calN'$, then, identifying $\calN \hat = \calN'$, permutation $\pi: \calN \to \calN, i \mapsto \tau(i)$ forms a player isomorphism from $\Gamma$ to $\Gamma'$. Indeed, action set sizes have to match since $\tau$ is a player symmetry. Therefore, it remains to show that for any $i \in \calN$ and $\bfa \in A$, we have
    \begin{align*}
        &(u')^{\pi(i)}(\bfa_{\pi^{-1}(1)}, \dots, \bfa_{\pi^{-1}(N)})
        \\
        &=(u')^{\tau(i)}(\bfa_{\tau^{-1}(1')}, \dots, \bfa_{\tau^{-1}(N')})
        \\
        &=\hat u^{\tau(i)}(\bot, \dots, \bot, \bfa_{\tau^{-1}(1')}, \dots, \bfa_{\tau^{-1}(N')})
        \\
        &\overset{(*)}{=}\hat u^{i}(\bfa_{1}, \dots, \bfa_{N}, \bot, \dots, \bot)
        \\
        &=u^i(\bfa_{1}, \dots, \bfa_{N}) \, ,
    \end{align*}
    where in $(*)$ we used that $\tau$ is a player symmetry of $\hat \Gamma$. 
    
    \paragraph{Observation 4:} If there exists a player symmetry $\tau$ of $\hat \Gamma$ mapping $\tau(\calN) = \calN'$, then any generator set of $\sym_{\hat \calN}(\hat \Gamma)$ must contain at least one such player symmetry, due to Observation 2.
    
    \paragraph{Conclusion:}
    In order to determine a player isomorphism $\pi \in \iso_\calN(\Gamma, \Gamma')$, if existent, compute a generator set of the automorphism group $\sym_{\hat \calN}(\hat \Gamma)$ via \Cref{thm:player symms}. This takes polytime in $|\Gamma|$ and $|\Gamma'|$ because $|\hat A| \leq \qty(\max\{|A|, |A'|\})^2$. If any generator $\tau$ satisfies $\tau(\calN) = \calN'$, use Observation 3 to obtain and return the desired player isomorphism $\pi \in \iso_\calN(\Gamma, \Gamma')$. If none satisfies $\tau(\calN) = \calN'$, then by Observation 4, there is no player symmetry $\tau'$ of $\hat \Gamma$ satisfying $\tau'(\calN) = \calN'$. By Observation 1, this means in particular that there cannot be a player isomorphism from $\Gamma$ to $\Gamma'$, and we can return that.
\end{proof}

Next, we discuss parameterized algorithms for computing general game symmetries and isomorphisms, allowing for ``nontrivial'' mappings between action sets. 

\allsymmcomp*
As argued in the main body, this implies the following.
\boundedactionspoly*

\begin{proof}[Proof of \Cref{thm:all symm algo}]
    We construct the same hypergraph as in \Cref{thm:player symms}, except now, we do not introduce a logsize encoding of action sets by subsets of nodes. This is because we are interested in game symmetries that map actions flexibly to actions. Therefore, we instead introduce one node $b_{ik}$ for \emph{each} action $a_k \in A^i$; we will also refer to it simply as $a_k$. In previous terms, we now have $k_i = m^i$. Note that the set of unique payoffs $P$ continue to be encoded as subsets of $U := \{ u_k \}_k$. In the third observation below, we investigate the complexity of constructing this graph. Before that, we will (again) analyze $\sym(\Gamma)$ through permutations of the vertices $V$ of this hypergraph. For that, associate any game symmetry $\phi \in \sym(\Gamma)$ with the vertex permutation $\psi_\phi : V \to V$ defined as $\calN \ni i \mapsto \phi(i)$, $A^i \ni a_k \mapsto \phi(a_k)$, and $u_k \mapsto u_k$. 
    
    \paragraph{Observation 1:} $\{ \psi_\phi : \phi \in \sym(\Gamma) \} = G \cap H$, where $G$ is the automorphism group of the hypergraph, and $H$ is the group of permutations that resemble game mappings, that is, all permutations $\psi : V \to V$ such that $\psi: \calN \overset{\text{bij}}{\to} \calN$, $\psi: A^i \overset{\text{bij}}{\to} A^{\psi(i)}$ for all $i \in \calN$, and $\psi|_P = \id_P$. Indeed, for ``$\supseteq$'' it is not hard to see that any $\psi \in G \cap H$ satisfies $\psi = \psi_\phi$ for a mapping $\phi$ defined as follows, forming a game symmetry: map $i \mapsto \psi(i)$ and $A^i \ni a_k \mapsto \psi(a_k)$.

    \paragraph{Observation 2:} Above observation implies that a generator set of $G \cap H$, restricted to $\calN \cup \acts \subset V$, forms a generator set of $\sym(\Gamma)$.

    \paragraph{Observation 3:} The hypergraph can be constructed in time $2^{\calO(|\acts|)}$. Let us prove that. It takes $\calO\qty((N \cdot |A|)^2)$ time to determine the set of unique payoffs $P$ and then construct the corresponding nodes $U$. Moreover, there are an additional $N + |\acts|$ nodes and an additional $N \cdot |A|$ hyperedges to be created. Hence, the hypergraph construction takes $\calO\qty(|A|^4)$ time. Notice that the quantity $|A|$ is largest for any given quantity $|\acts|$ exactly when we have the same action set sizes $m^1 = \dots = m^N = |\acts| / N$. Therefore, we derive 
    \begin{align*}
        \log_2 |A| \leq \log_2 \qty( \frac{|\acts|}{N} )^N = N \log_2 \frac{|\acts|}{N} \leq N \cdot \frac{|\acts|}{N} = |\acts| \, .
    \end{align*}
    This yields a running time
    \begin{align*}
        \calO\qty(|A|^4) &= \calO\qty( \qty(2^{\log_2 |A|})^4) = \calO\qty( \qty(2^{|\acts|})^4) = \calO\qty( 2^{4|\acts|} )
        \\
        &= 2^{\calO(|\acts|)} \, .
    \end{align*}
    
    \paragraph{Observation 4:} 
    The automorphism group $G$ can be computed in time $2^{\calO(|\acts|)}$. This is because of \Cref{th:hypergraph} since the hypergraph has
    \[
        |V| = N + |\acts| + \lceil \log_2 |P| \rceil \leq 2|\acts| + \lceil \log_2 |A| \rceil = \calO(|\acts|)
    \]
    many vertices by the estimation from Observation 3.
    
    \paragraph{Observation 5:} The permutation group $H$ is generated by the set of action swaps and the set of player swaps, that is, $\{ (b_{ik} b_{il}) \, \mid \, i \in \calN, 1 \leq k,l \leq m^i\}$ and $\{ (ij)(b_{i1} b_{j1}) \dots (b_{im^i} b_{jm^i}) \, \mid \, i,j \in \calN \text{ with } m^i = m^j \}$. We can list those generators in time $\calO(|\acts|^3 \cdot |V|) = 2^{\calO(|\acts|)}$.

    \paragraph{Conclusion} 
    Compute $G$ and $H$ as described in Observations 4 and 5, the intersection $G \cap H$ , and its intersection with \Cref{lem:coset intersection}. All of this takes $2^{\calO(|\acts|)}$ time. Observation 2 then yields $\sym(\Gamma)$.
\end{proof}
The following result is proven analogously to \Cref{lem:player iso coset or empty}.
\begin{lemma}
\label{lem:general iso coset or empty}
    For any two games $\Gamma, \Gamma'$, either $\iso(\Gamma, \Gamma')$ is empty, or it is a left coset $\iso(\Gamma, \Gamma') = \psi \sym(\Gamma)$ for $\psi \in \iso(\Gamma, \Gamma')$ chosen arbitrarily.
\end{lemma}
\begin{thm}\label{thm:iso comp}
    Given two games $\Gamma$ and $\Gamma'$, the coset of isomorphisms $\iso(\Gamma, \Gamma')$ is computable in time $2^{\calO(|\acts|)}$.
\end{thm}
\begin{proof}
    The proof is analogous to \Cref{thm:player iso}, except that we have to use \Cref{thm:all symm algo} and \Cref{lem:general iso coset or empty}, and that we have to allow and adapt for actions now being permuted as well. For example, to any game isomorphism $\psi \in \iso(\Gamma, \Gamma')$, we now have to associate a mapping $\phi_{\psi}$ for the constructed game $\hat \Gamma$ that still maps $\phi_{\psi}(i) = \psi(i)$ for $i \in \calN \subset \hat \calN$ and $\phi_{\psi}(a') = \psi^{-1}(a')$ for $a' \in A^{i'}$, but additionaly also maps $\phi_{\psi}(a) = \psi(a)$ for $a \in A^i$ and $\phi_{\psi}(i') = \psi^{-1}(i')$ for $i' \in \calN' \subset \hat \calN$. Below we adjust two parts of the proof to our new setting:
    
    \paragraph{Observation 1.5:} Prior to Observation 2, we now need to first observe that in any gamy symmetry $\psi$ of $\hat \Gamma$, an action $\bot$ must always be mapped to action $\bot$ again. To see that, assume there exists $i \in \hat \calN$, \wlogg $i \in \calN \subset \hat \calN$, with $\phi(\bot \in A^i) \neq \bot \in A^{\phi(i)}$. Then, because of $\phi$ being a symmetry and $m^i \geq 2$, we have that there must in return be two actions $\bar a, \bar a' \in A^i$ with $\phi(\bar a) = \bot \in A^{\phi(i)}$ and $\phi(\bar a') \neq \bot \in A^{\phi(i)}$. We note that in game $\hat \Gamma$, player $i$ receives a payoff $\neq u^\bot$ under profiles $\bar \bfa := (\bar a, \bfa_{-i}, \bot_{A'})$ and $\bar \bfa' := (\bar a', \bfa_{-i}, \bot_{A'})$ for any $\bfa_{-i} \in A^{-i}$, because $\bot(\bar \bfa) = \bot(\bar \bfa') = \calN'$. Because $\phi$ is a symmetry, this implies that player $\phi(i)$ must receive a payoff $\neq u^\bot$ under profile $\phi(\bar \bfa)$ and $\phi(\bar \bfa')$, which implies that $\bot ( \phi(\bar \bfa) ) = \bot ( \phi(\bar \bfa') ) \in \{\calN, \calN'\}$. However, this cannot be true since $\bar \bfa$ and $\bar \bfa'$ only differ in one entry whose image under $\phi$ is $=\bot$ and $\neq \bot$ respectively. Hence, a contradiction. Analogous reasoning also works for if $i$ was element in $\calN'$.

    \paragraph{Observation 3':} The derivations here, for game symmetry $\phi$ of $\hat \Gamma$ and associated isomorphism $\psi$ from $\Gamma$ to $\Gamma'$ will now look as follows for $i \in \calN$ and $\bfa \in A$:
    \begin{align*}
        &(u')^{\psi(i)}(\psi(\bfa)) =(u')^{\phi(i)}(\phi(\bfa)) = \hat u^{\phi(i)}(\bot, \dots, \bot, \phi(\bfa))
        \\
        &= \hat u^{\phi(i)}\big(\phi(\bfa, \bot, \dots, \bot)\big) = \hat u^{i}(\bfa, \bot, \dots, \bot) = u^{i}(\bfa) \, .
    \end{align*}
    
\end{proof}

\subsection{Search for Particular Properties}
\label{app:effisowithprops}

We can leverage the computation from \Cref{sec:effsymm} (\Cref{thm:all symm algo,thm:player symms,thm:player iso,thm:iso comp}) to further efficiently compute the symmetries and isomorphisms satisfying certain properties. Below, we merely give an example of this:
\begin{prop}
    Given game $\Gamma$ and, optionally, another game $\Gamma'$, we can compute the following groups / cosets:
    \begin{enumerate}
        \item In time $2^{\calO(|\acts|)}$ the game symmetries $\phi$ of $\Gamma$, or the game isomorphisms $\phi: \Gamma \to \Gamma'$, that map a particular action $a \in \acts$ to another action $\phi: a \mapsto b$, and
        \item In polytime the game symmetries $\phi$ of $\Gamma$, or the game isomorphisms $\phi: \Gamma \to \Gamma'$, that map a particular player $i \in \calN$ to another player $\phi: i \mapsto j$.
    \end{enumerate}
\end{prop}
\begin{proof}
    These results are all proved in a similar fashion, so let us consider the case of game isomorphisms $\phi: \Gamma \to \Gamma'$ mapping $a \mapsto b$. Call this coset $\calC$. As before, fix an ordering of players in $\Gamma$ and $\Gamma'$ such that we can identify the player sets with each other and have same respective action set sizes. (If that is not possible, $\Gamma$ and $\Gamma'$ cannot be isomorphic at all.) Then fix an order of the actions to identify the action sets with each other as well.
    
    Let ${\sf Stab}(a)) < \calS_{\acts}$ be the (stabilizing) subgroup of permutations $\pi : \acts \to \acts$ that fix $a$, \ie, $\pi(a) = a$. Moreover, let $\sigma_{ab}$ be the permutation $\acts \to \acts$ that swaps actions $a$ and $b$, and keeps the other actions fixed. Then it is not hard to see that
    \[
        \calC = \iso(\Gamma, \Gamma') \cap \qty(\sigma_{ab} \circ {\sf Stab}(a)) \, .
    \]
    ${\sf Stab}(a))$ is generated by all action swaps $(a',b')$ where $a' \neq a \neq b'$. With that, and \Cref{thm:iso comp} and \Cref{lem:coset intersection}, this yields a $2^{\calO(|\acts|)}$ time method to compute $\calC$.

    For the player isomorphisms case, we are now working with permutations on the player set $\calN$ and we instead need to observe that
    \[
        \calC' = \iso_\calN(\Gamma, \Gamma') \cap \qty(\sigma_{ij} \circ {\sf Stab}(i)) \, .
    \]
\end{proof}

\subsection{Excursion to Graphical Games}

\graphicalgames*
\begin{proof}
    Given a graph $G=(V,E)$, view it as a graphical game in which each node / player $v \in V$ has two actions $a_0,a_1$. First, we define player $v$'s payoffs for the corner cases when $v$ has no neighbors (=is an isolated node): Then, define its payoffs as $-1$ if that player plays $a_0$, and $-2$ if that player plays $a_1$. The purpose of these values is simply to be distinct from the payoffs a node with neighbors might receive. Namely, in that case, define $v$'s payoffs as
    $$
        \sum_{v'\text{ adjacent to }v} 1\left[ v \text{ and } v' \text{ play } a_1\right],
    $$
    where $1[P]$ is the indicator function evaluating as $1$ if property $P$ is satisfied, otherwise evaluating as $0$. 
    
    We can observe that a graph automorphism of $G$ forms a player symmetry of the graphical game (as defined above \Cref{thm:player symms}). We can also show a reverse direction of this statement, starting from a game symmetry $\phi$ of the graphical game. First, we note that the payoffs force $\pi$ to map isolated nodes to isolated nodes, and non-isolated nodes to non-isolated nodes. This, in turn, forces $\phi$ to map action $a_i$ to $a_i$. In particular, $\phi$ is a player symmetry. With this knowledge, we obtain that if $v$ and $v'$ are neighbors, then the action of $v'$ has an effect on the payoff of $v$ if $v$ plays $a_1$, hence we there is a same effect of $\pi(v')$ on $\pi(v)$, and therefore $\pi(v')$ and $\pi(v)$ are neighbors. Thus, $\pi$ of $\phi$ is a graph automorphism of $G$. 
    
    Analogous reasoning works for reducing the graph isomorphism problem for two graphs to the game isomorphism problem for two graphical games.
\end{proof}
\section{Further background on \Cref{sec:NE prelim}}

In this section, we give more context to \Cref{sec:comp symm} and prove statements from it.

First, we observe that any game symmetry naturally extends to strategy profiles: A strategy profile $\bfs$ is mapped to $\phi(\bfs)$ that plays an action $a \in A^j$ of player $j$ with the probability that $\bfs$ plays the associated action $\phi^{-1}(a)$ of associated player $\pi^{-1}(j)$. Satisfying the second property in \Cref{defn:game sym} (as it is phrased for action profiles $\bfa \in A$) is already enough to ensure that the analogous statement holds for the bigger set of all strategy profiles $\bfs \in S$, because in that case
\begin{align*}
    u^i(\bfs) &= \sum_{\bfa \in A} \bfs_1(\bfa_1) \cdot \ldots \cdot \bfs_N(\bfa_N) \cdot u^i(\bfa)
    \\
    &= \sum_{\bfa \in A} \phi(\bfs)_{\pi(1)}( \phi(\bfa_1 ) ) \cdot \ldots \cdot \phi(\bfs)_{\pi(N)}( \phi(\bfa_N ) )  
    \\
    &\quad \quad \quad \cdot u^{\pi(i)}(\phi(\bfa))
    \\
    &\overset{\bfa = \phi^{-1}(\bfa')}{=} \sum_{\bfa' \in A} \phi(\bfs)_{\pi(1)}( \phi(\phi^{-1}(\bfa')_1 ) ) \cdot \ldots 
    \\
    &\quad \quad \quad \cdot \phi(\bfs)_{\pi(N)}( \phi(\phi^{-1}(\bfa')_N ) ) \cdot u^{\pi(i)}(\bfa')
    \\
    &= \sum_{\bfa' \in A} \phi(\bfs)_{\pi(1)}( \bfa'_{\pi(1)} ) \cdot \ldots \cdot \phi(\bfs)_{\pi(N)}( \bfa'_{\pi(N)} ) 
    \\
    &\quad \quad \quad \cdot u^{\pi(i)}(\bfa')
    \\
    &\overset{\textnormal{reorder}}{=} \sum_{\bfa' \in A} \phi(\bfs)_1( \bfa'_1 ) \cdot \ldots \cdot \phi(\bfs)_N( \bfa'_N ) \cdot u^{\pi(i)}(\bfa')
    \\
    &= u^{\pi(i)}(\phi(\bfs)) \, .
\end{align*}

\symmpresnes*
\begin{proof}
    For action $a \in A^i$, we have 
    \begin{align}
    \label{eq:dev inc after symm}
        u^i (a, \bfs_{-i}) - u^i(\bfs) = u^{\pi(i)}(\phi(a), \phi(\bfs)_{-\pi(i)}) - u^{\pi(i)}(\phi(\bfs))
    \end{align}
    and therefore, there will be a player (say, $i$) with positive deviation incentives under $\bfs$ if and only if there is a player (then, $\pi(i)$) with positive deviation incentives under $\phi(\bfs)$. 
\end{proof}

\symmseqvltoorbits*

\begin{proof}
    If a profile respects symmetries $\phi$ and $\phi'$ in $\Sigma$, then it must also respect symmetry $\phi' \circ \phi$. Since $\sym(\Gamma)$ is finite, this suffices to say that respecting $\Sigma$ implies respecting its generated set $\gen{\Sigma}$. But strategy profile $\bfs$ respecting $\gen{\Sigma}$ means no more than $\bfs_i(a) = \bfs_{\pi(i)}(\phi(a))$ for all $\phi \in \gen{\Sigma}$ for all $a \in A^i$. This is equivalent to requiring $\bfs_i(a) = \bfs_{i'}(a')$ for all $a' \in A^{i'}$ ($i' \in \calN$) that is in the same orbit as $a \in A^i$, where $a \in \acts$ arbitrary. This yields the second statement.
\end{proof}

At this point, we might highlight the following two results from the literature for an interested reader: Since $\gen{\Sigma}$ and $\acts$ are finite, we get from the orbit-stabilizer theorem and Lagrange's theorem that the size $|\gen{\Sigma} a|$ of an orbit is exactly equal to the number of symmetries $|\gen{\Sigma}|$ divided by the number of symmetries $\phi \in \gen{\Sigma}$ that map $a$ to itself (``fixing $a$''). Thanks to Burnside's Lemma we have that the number of distinct orbits equals the average number of actions that a symmetry of $\gen{\Sigma}$ keeps fixed:
\[
    W(\gen{\Sigma}) = \frac{1}{|\gen{\Sigma}|} \sum_{\phi \in \gen{\Sigma}} |\{ a \in \acts : \phi(a) = a \}| \, .
\]

\explicittoorbits*

\begin{proof}
    We build the orbits orbit by orbit. Create a queue $Q$ consisting of all actions $\acts$. Take the next element $a$ from the queue $Q$. If it is not marked yet, create a new (yet uncompleted orbit) set $O = \{a\}$, mark $a$ as ``orbit already identified'', and add a pointer from it to $O$ that indicates that $O$ will be $a$'s orbit. Moreover, create an additional queue $Q_O$ for adding to $O$, initially consisting only of $\{a\}$. Enter this subloop that fully builds $O$, only after which we will go back to outer loop queue $Q$. The subloop: Take the next element $a'$ from queue $Q_O$ and loop through the list of symmetries. For each such symmetry $\phi$, check if $\phi(a')$ is already marked. If not, add it $O$, mark it, add a pointer from it to $O$, and add it to the queue $Q_O$. If it was already marked, then it must have already been in $O$ and we can ignore it. Once done looping through the symmetries, do the same looping again with the next action in queue $Q_O$, all up until $Q_O$ becomes empty. At that point, we have found all ways one may apply compositions of symmetries to $a$ to get to new actions in $\acts$. So $O$ is fully constructed, and we can go back to the outer loop and get the next element $a$. If it has not been marked yet, progress as described above. If it has been marked though, we can skip it because we have already identified its orbit in a previous encounter. Running time analysis: Each action is dealt with once, and when being dealt with at most once in queue $Q$ and once in a queue $Q_O$. The former process takes constantly many operations, the latter operation takes $\calO(|\Sigma|)$ many operations. Hence $\calO(|\Sigma| \cdot |\acts|)$.
\end{proof}

\section{On \Cref{sec:eq search}}

In this section, we give proofs to the results in \Cref{sec:eq search}.

\symppad*

\begin{proof}
    \PPAD{}-hardness is straight-forward: In cases where $\Sigma = \{\id_{\acts}\}$, we are facing the problem of computing any $\eps$-\NE{} of the game, which is \PPAD{}-complete \cite{DaskalakisGP09,ChenDT09}. We also obtain this result for a nontrivial set of symmetries through a well-known trick that is related to not disclosing to participants which player identity they might play as \cite{GaleKT52}: \NEs{} of a bimatrix game $(A,B)$---which we can \wlogg assume to only have positive entries by shifting payoffs---correspond to player-symmetric \NEs{} of the totally symmetric bimatrix game  defined by $M := \begin{pmatrix} 0 & A \\ B^T & 0 \end{pmatrix}$, see \citet{NisanTV07}[Theorem~2.4].

    We develop \PPAD{}-membership using the general framework of \citet{EtessamiY10} for showing that finding a fixed point of a Brouwer function $F: D \to D$ is in \PPAD{}. Our domain $D$ will be the set of strategy profiles respecting the symmetries. Our Brouwer function will be Nash's function \cite{Nash51}, which was defined for the larger set $S$ of all strategy profiles: Given a profile $\bfs \in S$, player $i \in \calN$, and action $a \in A^i$, define the advantage of $a$ (over $\bfs_i$) as
    \[
        g(\bfs,i,a) := u^i (a,  \bfs_{-i}) - u^i (\bfs)
    \]
    and profile $F(\bfs)$ as 
    \[
        F(\bfs)_i(a) := \frac{ \bfs_i( a ) + \max\{0,g(\bfs,i,a)\} }{ 1 + \sum_{a' \in A^i} \max\{0,g(\bfs,i,a) \} } \, .
    \]
    \citeauthor{Nash51} shows that $F(\bfs)$ for $\bfs \in S$ indeed makes a profile in $S$ again, \ie, $F : S \to S$. He also gives a short and high-level argument for why $F$ maps symmetries-respecting profiles to symmetries-respecting profiles, that is, $F|_D : D \to D$. For completeness, we work out the details of this observation at the end of this proof. \citeauthor{EtessamiY10} moreover show that (1) the functions $F$ constructed from a game $\Gamma$ are nicely Lipschitz continuous and therefore satisfy the requirement of being \emph{polynomially continuous}, and that (2) an $\eps'$-fixed point $\bfs \in S$ of $F$ makes an $\eps$-\NE{} of the  $\Gamma$ for some $\eps$ in $\poly$ encoding size of the encoding of $\Gamma$ and $\eps'$. Naturally, the functions $F_D$ inherit both of these properties from the functions $F$. It remains to show that $D$ is a convex polytope efficiently describable by linear (in-)equalities; because then, we have that functions $F_D$ are also \emph{polynomially computable}, implying the \PPAD{} result by invoking \cite{EtessamiY10}[Proposition 2.2]. 
    
    \paragraph{$D$ is easy to describe:} First, get the symmetries in orbit form $W$ via \Cref{lem:explicit to orbit} (if not already given in that form). Next, choose one representative $a_{\omega}$ of each orbit $\omega \in W$. Then, we can describe $D$ through the following linear (in-)equality system:
    \begin{align}
    \label{D as LES}
    \begin{aligned}
        &\bfs_i(a) \geq 0 &&\forall i \in \calN, a \in A^i
        \\
        & \sum_{a \in A^i} \bfs_i(a) = 1 &&\forall i \in \calN
        \\
        &\bfs_{i(a)}(a) = \bfs_{i(a_{\omega})}(a_{\omega}) &&\forall \omega \in W, a \in \omega \setminus \{a_{\omega}\} \ \, ,
    \end{aligned}
    \end{align}
    where we use the notation $i(a')$ to denote the player $i$ with $a' \in A^i$. These $2|\acts| + N$ many (in-)equalities together with their coefficients can be computed in polytime in the encoding of $\Gamma$ and $W$.
    
    \paragraph{$F$ maps $D$ to $D$:} Let $\bfs \in D$, $i \in \calN$, and $a \in A^i$. Then, get $\omega \in W$ such that $a \in w$, and a symmetry $\phi$ with $\phi(a) = a_{\omega} \in A^{i(a_{\omega})}$ such that there exists $\phi \in \Sigma \subseteq \sym(\Gamma)$ with $W = W(\gen{\Sigma})$. We have to show that $F(\bfs)_{i(a_{\omega})}(a_{\omega}) = F(\bfs)_{i}(a)$. Then $F(\bfs)$ satisfies the system in (\ref{D as LES}), yielding membership in $D$. To that end, we observe for all $a' \in A^{i}$ that 
    \begin{align*}
        g(\bfs, i, a') &\overset{(\ref{eq:dev inc after symm})}{=} g(\phi(\bfs), \pi(i), \phi(a')) 
        \\
        &\overset{\bfs \in D}{=} g(\bfs, i(a_{\omega}), \phi(a')) \, .       
    \end{align*}
    In particular, we have
    \[
        g(\bfs, i, a) = g(\bfs, i(a_{\omega}), a_{\omega}) \, ,
    \]
    and
    \begin{align*}
        &\sum_{a' \in A^{i}} \max \{ 0, g(\bfs, i, a') \} 
        \\
        &= \sum_{a' \in A^{i}} \max \{ 0, g(\bfs, i(a_{\omega}), \phi(a'))  \} 
        \\
        &= \sum_{\tilde a \in A^{i(a_{\omega})}} \max \{ 0, g(\bfs, i(a_{\omega}), \tilde a)  \} \, .
    \end{align*}
    Therefore, we can combine these observations to conclude that indeed we have
    \begin{align*}
        &F(\bfs)_{i(a_{\omega})}(a_{\omega}) = \frac{ \bfs_{i(a_{\omega})}( a_{\omega} ) + \max\{0,g(\bfs,i(a_{\omega}),a_{\omega})\} }{ 1 + \sum_{\tilde a \in A^{i(a_{\omega})}} \max\{0,g(\bfs,i(a_{\omega}),\tilde a) \} }
        \\
        &= \frac{ \bfs_i( a ) + \max\{0,g(\bfs,i,a)\} }{ 1 + \sum_{a' \in A^i} \max\{0,g(\bfs,i,a') \} } = F(\bfs)_{i}(a) \, .
    \end{align*}
\end{proof}

\symcls*

\begin{proof} \,
    We will begin with proving \CLS{}-membership.

    \paragraph{Reduction to KKT} We show that {\sc Sym-Nash-Team} reduces to the \CLS{}-complete problem of finding a KKT point of a nice enough function over a convex compact polytope. We assume familiarity with the definition of that problem {\sc KKT} here, which can be found in \cite{FearnleyGHS23}[Page 21]. Let a {\sc Sym-Nash-Team} instance be given, that is, a team game $\Gamma$ in explicit form, a precision parameter $\eps > 0$, and symmetries $W$ of $\Gamma$ in orbit form.
    
    Analogous to the proof of \Cref{thm:PPAD}, choose one representative $a_{\omega}$ of each orbit $\omega \in W$, and let $D$ be the non-empty convex compact polytope as described in (\ref{D as LES}) that captures the set of strategy profiles that respect the given symmetries. Denoting with $w(a) \in W$ the orbit of an action $a \in A^i$, we can rewrite the description of $D$ to:
    \begin{align*}
    % \label{D as LES}
    \begin{aligned}
        &\bfs_i(a) \geq 0 &&\forall i \in \calN, a \in A^i
        \\
        & \sum_{a \in A^i} \bfs_i(a) = 1 &&\forall i \in \calN
        \\
        &\bfs_{i}(a) = \bfs_{i(a_{\omega(a)})}(a_{\omega(a)}) &&\forall i \in \calN, a \in A^i \text{ s.t. } a \neq a_{\omega(a)} \, .
    \end{aligned}
    \end{align*}
    
    Each equality constraint can be converted into two inequality constraints. The team's utility function $u$ and its partial derivatives are the following multivariate polynomial functions:
    \[
        u(\bfs) = \sum_{\bfa \in A} u(\bfa) \prod_{i \in \calN} \bfs_i(\bfa_i)
    \]
    and for $i \in \calN, a \in A^i$:
    \[
        \nabla_{i,a} \, u(\bfs) = \sum_{\bfa^- \in A^{-i}} u(a, \bfa^-) \prod_{j \neq i} \bfs_j(\bfa_j^-)
    \]
    for $i \in \calN$ and $a \in A^i$. They can be constructed in polytime, with monomial coefficients derived from the payoffs $\{u(\bfa)\}_{\bfa}$. A Lipschitz constant $L > 0$ that shows that both $u$ and $\nabla u$ are Lipschitz continuous over $D$ (via derivatives of $u$ and $\nabla u$) can also be constructed in polytime. Both functions---as simple-to-describe polynomials---make \emph{well-behaved arithmetic circuits}. Finally, choose the precision parameter of {\sc KKT} as $\delta := \eps / 2 > 0$. Then, we satisfy the conditions of the computational problem {\sc KKT}. Since the gradient and Lipschitz constant are computed correctly, the problem returns a $\delta$-KKT point of the optimization problem of $u$ over $D$. A $\delta$-KKT point $\bfs \in \R^{|A|}$ for our problem is characterized by the following conditions: there exist KKT-multipliers $(\mu_{i,a})_{i \in \calN, a \in A^i}$, $(\kappa_i)_{i \in \calN}$, and $(\tau_{i,a})_{i \in \calN, a \in A^i : a \neq a_{\omega(a)}}$ in $\R$ such that
    \begin{align*}
        &\bfs_i(a) \geq 0 &&\forall i \in \calN, a \in A^i
        \\
        & \sum_{a \in A^i} \bfs_i(a) = 1 &&\forall i \in \calN
        \\
        &\bfs_{i}(a) = \bfs_{i(a_{\omega(a)})}(a_{\omega(a)}) &&\forall i \in \calN, \forall a \in A^i : a \neq a_{\omega(a)}
        \\
        &\mu_{i,a} \geq 0 &&\forall i \in \calN, \forall a \in A^i
        \\
        &\mu_{i,a} = 0 \quad \textnormal{or} \quad \bfs_i(a) = 0 &&\forall i \in \calN, \forall a \in A^i
    \end{align*}
    \begin{flalign}
    \label{eq:KKT obj conds}
    \begin{aligned}
        &| \nabla_{i,a} \, u(\bfs) + \mu_{i,a} - \kappa_i - \tau_{i,a}) | \leq \delta&& 
        \\
        &\, \quad \quad \quad \quad \quad \quad \quad \quad \forall i \in \calN, \forall a \in A^i : a \neq a_{\omega(a)}&& 
        \\
        &| \nabla_{i,a} \, u(\bfs) + \mu_{i,a} - \kappa_i + \sum_{a' \in w(a) \setminus \{a\}} \tau_{i(a'),a'}) | \leq \delta&& 
        \\
        &\, \quad \quad \quad \quad \quad \quad \quad \quad \forall i \in \calN, \forall a \in A^i : a = a_{\omega(a)} \, .&& 
    \end{aligned}
    \end{flalign}

    We will show that $\bfs$ must then be an $\eps$-\NE{} of $\Gamma$. 
    
    \paragraph{Well-Supportedness} In fact, we show a stronger property which is that $\bfs$ is an $\eps$-well-supported \NE{}. This is the case if $\forall i \in \calN, a \in A^i$ with $\bfs_i(a)>0$, we have that for all $\tilde a \in A^i$:
    \[
        u(a,\bfs_{-i}) \geq u(\tilde a,\bfs_{-i}) - \eps \, .
    \]
    As a first step, we will argue now that in order to check the above condition, because of $\bfs \in D$, it suffices to check for all $\omega \in W$ with $\bfs_{i_\omega}(a_w)>0$, that for all $\tilde a \in A^{i_\omega}$
    \begin{align}
    \label{eq:orbit NE check}
        u(a_\omega,\bfs_{-i_\omega}) \geq u(\tilde a,\bfs_{-i_\omega}) - \eps \, .        
    \end{align}
    And indeed, with any symmetry $\phi : a \mapsto a_\omega$ that is element of some $\Sigma \subseteq \sym(\Gamma)$ with $W = W(\gen{\Sigma})$, such a $\phi$ implies for any $\bfs \in D$ that
    \begin{enumerate}
        \item $\bfs_i(a)>0 \iff \bfs_{i_\omega}(a_\omega)>0$,
        \item 
            \begin{align}
                \label{eq:phi and grad}
                \nabla_{i(a),a} \, u(\bfs) = u(a, \bfs_{-i}) = u(a_\omega, \bfs_{-i_\omega}) = \nabla_{i_\omega,a_\omega} \, u(\bfs) \, ,
            \end{align}
        \item $u(a, \bfs_{-i}) - u(\tilde a, \bfs_{-i}) = u(a_\omega, \bfs_{-i_\omega}) - u(\phi(\tilde a),\bfs_{-i_\omega})$.
    \end{enumerate}
    An advantage of well-supportedness is the following: For any orbit $\omega$ with $\bfs_{i_\omega}(a_w)>0$, we also have $\bfs_i(a)>0$ for all $a \in \omega$. By the KKT-conditions, this implies 
    \begin{align}
    \label{eq:mus sum to zero}
        \sum_{a \in \omega} \mu_{i,a} = \sum_{a \in \omega} 0 = 0 \, .
    \end{align}
    
    \paragraph{Cardinality} 
    Compute the orbit cardinalities $c(\omega) = |A^{i_{\omega}} \cap \omega| > 0$ for all $\omega \in W$. This takes $\calO(|\acts| \cdot |W|)$ time to compute. The cardinality can be interpreted as the number of actions an orbit contains from a(ny) single player $i$ with $\omega \cap A^i \neq \emptyset$, and it is indeed the same, independent of the chosen representative $a_{\omega}$: Suppose $a \in \omega \cap A^i$ and $a' \in \omega \cap A^{i'}$ are two possible representatives. If $i = i'$, then we immediately obtain the same cardinality. If $i \neq i'$, then a game symmetry $\phi$ that maps $a \mapsto a'$ and for which there exists $\phi \in \Sigma \subseteq \sym(\Gamma)$ with $W = W(\gen{\Sigma})$, bijectively associates any $b \in \omega \cap A^i$ with a $b' = \phi(b) \in \omega \cap A^{i'}$.

    \paragraph{Simplifying}
    First, we can combine the two $\delta$-inequalities of (\ref{eq:KKT obj conds}) to obtain for each orbit $\omega \in W$ that
    \begin{align}
        &\Big| |\omega| \cdot \nabla_{i_\omega,a_\omega} \, u(\bfs) + \sum_{a \in \omega} \mu_{i(a),a}  - \sum_{a \in \omega} \kappa_{i(a)} \Big| \label{eq:mu and kappa}
        \\
        &\overset{(\ref{eq:phi and grad})}{=} \Big| \nabla_{i_\omega,a_\omega} \, u(\bfs) + \mu_{i_\omega,a_\omega} - \kappa_{i_\omega} + \sum_{a \in \omega \setminus \{ a_\omega \}} \tau_{i(a),a} \nonumber
        \\
        &\, \quad + \sum_{a \in \omega \setminus \{ a_\omega \}} \qty( \nabla_{i(a),a} \, u(\bfs) + \mu_{i(a),a}  - \kappa_{i(a)} - \tau_{i(a),a} ) \Big| \nonumber
        \\
        &\overset{\Delta\text{-Ineq.}}{=} \Big| \nabla_{i_\omega,a_\omega} \, u(\bfs) + \mu_{i_\omega,a_\omega} - \kappa_{i_\omega} + \sum_{a \in \omega \setminus \{ a_\omega \}} \tau_{i(a),a} \Big| \nonumber
        \\
        &\, \quad + \sum_{a \in \omega \setminus \{ a_\omega \}} \Big| \qty( \nabla_{i(a),a} \, u(\bfs) + \mu_{i(a),a}  - \kappa_{i(a)} - \tau_{i(a),a} ) \Big| \nonumber
        \\
        &\overset{(\ref{eq:KKT obj conds})}{\leq} \delta + \sum_{a \in \omega \setminus \{ a_\omega \}} \delta = |\omega| \cdot \delta \, . \nonumber
    \end{align}
    For orbits $\omega \in W$ with $\bfs_{i_\omega}(a_w)>0$, this, in particular, implies
    \begin{align}
        \label{eq:grad to kappa}
        \begin{aligned}
            &\nabla_{i_\omega,a_\omega} \, u(\bfs) = \frac{1}{|\omega|} \cdot \qty( |\omega| \cdot \nabla_{i_\omega,a_\omega} u(\bfs) + \sum_{a \in \omega} \mu_{i(a),a} )
            \\
            &\overset{(\ref{eq:mus sum to zero})}{\geq} \frac{1}{|\omega|} \qty( \sum_{a \in \omega} \kappa_{i(a)} - |\omega| \delta ) = \frac{1}{|\omega|} \sum_{a \in \omega} \kappa_{i(a)} - \delta \, .
        \end{aligned}
    \end{align}
    Secondly, consider another action $a_\omega \neq \tilde a \in A^{i_\omega}$, and denote its orbit with $\tilde \omega := \omega(\tilde a)$. Then, we can further show that
    \begin{align}
        \label{eq:kappas for w's}
        \frac{1}{|\omega|} \sum_{a \in \omega} \kappa_{i(a)} = \frac{1}{|\tilde \omega|} \sum_{a \in \tilde \omega} \kappa_{i(a)} \, .
    \end{align}
    To see this, notice that the $\kappa_{i(a)}$'s only depend on the player index $i$. Define $\calI(\omega) = \{i \in \calN: \omega \cap A^i \neq \emptyset\}$ as the player set that $\omega$ intersects, and $\calI(\tilde \omega)$ analogously. Then, $i_\omega \in \calI(\omega) \cap \calI(\tilde \omega) $. This non-emptiness, in fact, already implies $\calI(\omega) = \calI(\tilde \omega)$: Any symmetry $\phi : a_\omega \mapsto a' \in A^{i'}$ for some $i' \in \calN$ implies $\phi(\tilde a) \in A^{i'}$ as well. Therefore, we can derive
    \begin{align*}
        &\frac{1}{|\omega|} \sum_{a \in \omega} \kappa_{i(a)} = \frac{c(\omega)}{|\omega|} \sum_{i \in \calI(\omega)} \kappa_i = \frac{1}{|\calI(\omega)|} \sum_{i \in \calI(\omega)} \kappa_i 
        \\
        &= \frac{1}{|\calI(\tilde \omega)|} \sum_{i \in \calI(\tilde \omega)} \kappa_i = \frac{1}{|\tilde \omega|} \sum_{a \in \tilde \omega} \kappa_{i(a)} \, .
    \end{align*}
    
    \paragraph{Concluding Membership} Finally, let us show that $\bfs$ is an $\eps$-well-supported \NE{}. As argued earlier, we have to show the following: For $\omega \in W$ with $\bfs_{i_\omega}(a_w)>0$ and $\tilde a \in A^{i_\omega}$, we have
    \begin{align*}
        &u(a_\omega,\bfs_{-i_\omega}) = \nabla_{i_\omega,a_\omega} \, u(\bfs) 
        \\
        &\overset{(\ref{eq:grad to kappa})}{\geq} \frac{1}{|\omega|} \sum_{a \in \omega} \kappa_{i(a)} - \delta\overset{(\ref{eq:kappas for w's})}{=} \frac{1}{|\tilde \omega|} \sum_{a \in \tilde \omega} \kappa_{i(a)} - \delta 
        \\
        &\overset{\mu \geq 0}{\geq} \frac{1}{|\tilde \omega|} \qty( - \sum_{a \in \tilde \omega} \mu_{i(a),a} + \sum_{a \in \tilde \omega} \kappa_{i(a)} ) - \delta  
        \\
        &\overset{(\ref{eq:mu and kappa})}{\geq} \frac{1}{|\tilde \omega|} \Big( |\tilde \omega| \cdot \nabla_{i_{\tilde \omega},a_{\tilde \omega}} \, u(\bfs) - |\tilde \omega| \delta \Big) - \delta  
        \\
        &= \nabla_{i_{\tilde \omega},a_{\tilde \omega}} \, u(\bfs) - 2\delta \overset{(\ref{eq:phi and grad})}{=} \nabla_{i_\omega,\tilde a} \, u(\bfs) - \eps
        \\
        &= u(\tilde a,\bfs_{-i_\omega}) - \eps       
    \end{align*}
    by the choice of $\delta$.

    \paragraph{Hardness} We refer to \citet{GhoshH24}'s concurrent work, which---as mentioned in the main body---develops a stronger \CLS{}-hardness result than us. Arguably, their proof approach is more fundamental than ours. More precisely, they leverage that the reverse connection holds to what we have shown in the membership proof, namely, that the search for a KKT point in a quadratic optimization problem over the simplex can be reduced to finding a \NE{} of a two-player game that is totally symmetric. Nonetheless, we will describe our proof approach for totally symmetric $5$-player games (or more players) below.

    We reduce hardness from a \CLS{}-completeness result by \citet{TewoldeOCG23,Tewolde+24} for single-player \emph{imperfect-recall games}. Those are extensive-form decision problems (\ie, single-player games), where the player's decision nodes are arbitrarily partitioned into infosets. Then it might be the case that the player forgets information at some decision node that it held at a previous decision node. \citet{LambertMS19} show that a particular kind of equilibrium (\emph{Causal Decision Theory} (CDT) equilibrium) always exists in such games by showing that they correspond to \NEs{} that respect a particular set of player symmetries in an associated perfect-recall game. We will argue below that with careful consideration, this construction gives rise to a polytime reduction for the computational search problems of these equilibria. Upon establishing that, we can combine it with \citet{TewoldeOCG23}[Theorem~2], which shows that finding a CDT equilibrium is \CLS{}-complete, even in single-player imperfect-recall games $G$ with a highly regular game tree: it has a constant depth of $6$ for every root-leaf paths, every nonterminal node (\ie, not leaf) is a decision node (\ie, not a chance node), belonging to the same single infoset of the game, having $m$ actions. In words, the player has to has to take one of $m$ actions $5$ times in a row without remembering whether it has taken a decision before or not (or what decision it took).
    
    So let us revisit the construction of the associated perfect-recall game $\Gamma$ to $G$ of the structure described above. Since the infoset in $G$ is visited $5$ times in a row by the player, we introduce a team of $5$ players PL1 to PL5 to $\Gamma$. Game $\Gamma$ then first uniformly randomly draws an ordering $\sigma$ among those $5$ players, without revealing the ordering to the players. Next, the players are asked one after the other to choose one of $m$ actions, without knowing whether other players have acted before them, or what action they chose. Note that we can freely permute the player identities in this game without any affects on the game, because of the uniform random draw of the player order at game start. In this paper's formalism, we can make this precise by looking at the normal-form representation of $\Gamma$ and observing that it is totally symmetric. What \citeauthor{LambertMS19} then showed is that CDT equilibria in $G$ correspond to \NEs{} of $\Gamma$ that respect this total symmetry, \ie, where all players play the same probability distribution over their $m$ actions. Since we are interested in approximation, they show that the Nash deviation incentives in $\Gamma$ are equal to a multiple of the CDT deviation incentives in $G$. For us, it is a multiple of $5$; in general, the multiple equals the expected frequency with which the infoset is visited. Hence, an $\frac{\eps}{5}$-\NE{} of $\Gamma$ that respect its total symmetry makes an $\eps$-CDT equilibria of $G$. 
    
    The construction is polytime: $\Gamma$ is $5!$ times as big as $G$, and $\Gamma$'s normal-form game representation is a payoff tensor of size $m^5$. Lastly, we enumerate the $5!$ (player) symmetries in explicit form.
\end{proof}

\lownumorbits*

\begin{proof}
    We show how to perform the equilibrium computation on the orbits instead of strategy profiles. Let $D$ be the set of strategy profiles that respect the given symmetries, which we may assume are in orbit form. For each orbit $\omega \in W$, let $a(\omega) \in \omega$ be a designated action representative of $\omega$ and $i(\omega)$ the player that $a(\omega)$ belongs to. Also compute the orbit cardinalities $c(\omega) = |A^{i(\omega)} \cap \omega| > 0$  for all $\omega \in W$, as in the proof of \Cref{thm:CLS}.
    
    We define the orbit profile set $E$ as any $\bfr \in \R^{|W|}$ that satisfies the following linear (in-)equality systen:
    \begin{align}
    \label{eq:orbit profiles char}
    \begin{aligned}
        &\bfr_{\omega} \geq 0 &&\forall \omega \in W
        \\
        &\bfr_{\omega} \leq 1 &&\forall \omega \in W
        \\
        &\sum_{w' \in W \text{ with } w' \cap A^{i(\omega)} \neq \emptyset} \bfr_{\omega'} = 1 &&\forall \omega \in W \, .
    \end{aligned}
    \end{align}

    We show that there is a one-to-one correspondance between $D$ and $E$, that is, between strategy profiles $\bfs$ that respect the orbits and orbit profiles. Given $\bfs \in D$ define $\bfr = \tilde \bfr(\bfs)$ as 
    \[
        \bfr_{\omega} := \sum_{a \in A^{i(\omega)} \cap \omega} \bfs_{i(\omega)} (a) = c(\omega) \cdot \bfs_{i(\omega)}(a(\omega)) \, .
    \]
    This works because since $\bfs$ respects the orbits, the sum above will be the same independent of the choice of representative $a(\omega)$ and, in fact, each summand will be equal, which also yields the latter equality. In particular, $\tilde \bfr(\bfs)$ satisfies (\ref{eq:orbit profiles char}). To see why the last equality holds, for each $\omega \in W$, we can imagine that each of these $\omega'$ have had their representitives chosen as $i(\omega') = i(\omega)$, which simplifies the sum to $\sum_{a \in A^{i(\omega)}} \bfs_{i(\omega)}(a) = 1$. Next, given an orbit profile $\bfr \in E$, define $\bfs = \tilde \bfs(\bfr)$ as $\bfs_i(a) = \frac{\bfr_{\omega}}{c(\omega)}$ where $\omega$ satisfies $a \in \omega$. By similar reasoning, $\bfs$ makes a strategy profile. It also respects the orbits because each action of the same orbit is assigned the same probability. We have that those two maps $\tilde \bfr : D \to E$ and $\tilde \bfs : E \to D$ are each other inverse (both-sided). Moreover, with precomputed representatives $a(\omega)$, cardinalities $c(\omega)$, and pointers that tell us to what orbit an action $a \in A^i$ belongs, we can compute $\tilde \bfs(\bfr)$ and $\tilde \bfr(\bfs)$ in linear time. 
    
    Now, construct all polynomial function $u^i : S \to \R$ in linear time. Next, construct utility functions $v^i := u^i \circ \tilde \bfs$. We remark a difference between $u^i$ and $v^i$: function $u^i$ is \emph{multilinear} in each $S^i$, \ie, in each monomial in $u^i$ at most one action probability $\bfs_i(a)$ occurs out of $\{ \bfs_i(a') \}_{a' \in A^i}$ for each player $i$. Functions $v^i$, on the other hand, are not, for example, they may have a $\bfr_{\omega}^k$ occur in a monomial with a higher degree $k \geq 2$ if $\omega$ covers actions from different players. Next, construct for each orbit $\omega \in W$ the deviation incentive function 
    \[
        g_{\omega}(\bfr) := u^{i(\omega)}(a(\omega), \tilde \bfs (\bfr)_{-i(\omega)}) \, ,
    \]
    which again is a polynomial function over $E$. With all of this, consider the additional inequalities:
    \begin{align}
    \label{eq:NE cond for orbit profiles}
        g_{\omega}(\bfr) \leq v^{i(\omega)}(\bfr) \quad \forall \omega \in W
    \end{align}

    Then we claim that symmetries-respecting $\bfs \in D$ is a \NE{} if and only if its associated $\bfr = \tilde \bfr (\bfs) \in E$ satisfies (\ref{eq:NE cond for orbit profiles}). This is because for any $a \in A^i$ we have (if $\omega$ denotes the orbit it belongs to):
    \begin{align*}
        u^i (a, \bfs_{-i}) - u^i(\bfs) &\overset{(\ref{eq:dev inc after symm})}{=} u^{i(\omega)}(a(\omega), \bfs_{-i(\omega)}) - u^{i(\omega)}(\bfs)
        \\
        &= g_{\omega}(\bfr) - v^{i(\omega)}(\bfr) \, .
    \end{align*}
    Thus, the former is nonpositive if and only if the latter is nonpositive.

    Therefore, we reduced in polytime the task of computing a symmetries-respecting \NE{} $\bfs$ to the task of finding an element in $\R^{|W|}$ that satisfies the joint system of (\ref{eq:orbit profiles char}) and (\ref{eq:NE cond for orbit profiles}). This system consists of $4|W|$ polynomial (in-)equalities, with the highest degree of an occuring polynomial being that of $v^i$ which is at most $N$ (for when an orbit covers at least one action of each player). The algorithm by \citet{renegar1992computational} for solving such a system takes polytime in all parameters of such a system except in $( \, \#\text{(in-)equalities } \cdot \, \#\text{degree} \, )^{\#\text{(in-)equalities }}$. We can turn this into an inverse-exponential approximation of a solution in $E$ with a subdivision algorithm on $E$ with query access to \citeauthor{renegar1992computational} (for details, compare with \citet{Tewolde+24}[Proposition 9]). Since the polynomial functions are nicely Lipschitz-continuous, inverse-exponential approximation in $E$ also implies inverse-exponentially decreasing deviation incentives $\eps$ in the definition of an $\eps$-\NE{}. All in all, that makes a $\poly\qty( |\Gamma|, \log(1/\epsilon), (|W|N )^{|W|} )$ to compute an orbit profile $\bfr$ whose associated strategy profile $\bfs = \tilde \bfs (\bfr)$ makes an $\eps$-\NE{} of $\Gamma$ that respects the given symmetries.
    
    Lastly, if we only consider the case of two players ($N=2$), then we can do better: use the {\em support enumeration} approach \cite{dickhaut1993program} instead, but in the orbit profile space $E$. For that, we note that the viable supports for a symmetries-respecting strategy profile $\bfs \in D$ are exactly all possible supports of associated orbit profile $\bfr = \tilde \bfr (\bfs)$. There are $2^{|W|}$ of such supports, and checking whether there exists an orbit profile with a given support that is also associated to a \NE{} $\bfs \in D$ can be done in polytime as usual.
\end{proof}

\twoplzerosumeasy*

For this result we will first prove a intermediate result:

\begin{lemma}
\label{lemma:rho regularizer}
    Let $\Gamma$ be a two-player zero-sum game. Let $f : [0, 1] \to \R$ be any strictly convex function, and define the regularizer $\rho: S^1 \times S^2 \to \R$ by $$\rho(\bfs) = \sum_{a \in \acts} f(\bfs(a)).$$
    Then the \NE{} $\bfs^* \in S$ that minimizes the regularizer $\rho$ also respects all symmetries.
\end{lemma}
\begin{proof}
    Let $\phi : \acts \to \acts$ be a symmetry of $\Gamma$. Note that $\phi$ is a permutation, and let $n$ be the order of $\phi$. That is, $n$ is the smallest positive integer such that $\phi^n$ is the identity. Define the sequence of strategies $\bfs^1 = \phi(\bfs^*), \bfs^2 = \phi^2(\bfs^*), \dots, \bfs^n = \phi^n(\bfs^*) = \bfs^*$. By \Cref{lem:symm preserves NE} each $\bfs^k$ is itself a Nash equilibrium of $\Gamma$. Since the set of Nash equilibria of any zero-sum game is convex, the average strategy profile $\bar\bfs := \frac{1}{n}\sum_{k=1}^n \bfs^k$ is also a Nash equilibrium. Moreover, $\rho$ is a function that respects symmetries in the sense that $\rho(\bfs) = \rho(\phi(\bfs))$, because $\phi$ only reorders the probability values in $\bfs$ and $\rho$ only depends on the present values. Therefore, since $\rho$ is also strictly concave, we have
    \begin{align*}
        \rho(\bar\bfs) \le \frac{1}{n} \sum_{k=1}^n \rho(\bfs^k) = \rho(\bfs^*)
    \end{align*}
    with equality only when $\bar\bfs = \bfs^k$ for all $k$. Since $\bfs^*$ minimizes $\rho$ among all Nash equilibria, we thus have that all the $\bar\bfs^k$'s are the same, \ie, $\bfs^*$ obeys the symmetry $\phi$. As symmetry $\phi$ was chosen arbitrarily, $\bfs^*$ obeys all symmetries of $\Gamma$.
\end{proof}
\Cref{thm:2p0s NEs easy} now follows by noting that finding a $\rho$-minimizing \NE{} can be expressed as a convex quadratic program.

\begin{proof}[Proof of \Cref{thm:2p0s NEs easy}]
    Let $\Gamma$ be a two-player zero-sum game described by P1's payoff matrix $U \in \R^{m_1 \times m_2}$. Thanks to the minimax theorem \cite{Neumann28}, $\Gamma$ then has a well-defined \emph{value} $v \in \R$, and its \NE{} set can be determined via the following linear (in-)equality system
    \begin{align*}
        &\bfs_1 \in S^1, \bfs_2 \in S^2
        \\
        &(U \bfs_2)_a \leq v \quad \, \forall a \in A^1
        \\
        &(\bfs_1^T U)_a \leq v \quad \forall a \in A^2 \, .
    \end{align*}
    Now add the quadratic objective function $\rho(\bfs_1, \bfs_2)$ to it that is defined as in \Cref{lemma:rho regularizer} by, for example, $f(x) := x^2$. This makes it a convex quadratic program which can be solved in polytime \cite{Kozlov80}.
\end{proof}

\end{document}